\documentclass[%
reprint,
amsmath,amssymb,
raggedbottom,
aps,
pra,
prb,
rmp,
floatfix,
showkeys,
prl, 
superscriptaddress
]{revtex4-1}

    \usepackage{mlmodern}

    \usepackage{hyperref}
    \hypersetup{colorlinks=true,
                allcolors=black,   
                filecolor=blue,
                citecolor=blue,
                linkcolor=blue,
                urlcolor=blue,
                menucolor=black,
                anchorcolor=black,
                }
	\usepackage{float}
	\usepackage{lipsum}
\usepackage{amsmath}
\usepackage{amssymb}
\usepackage{multirow}
\usepackage{graphicx}
\usepackage{dcolumn}
\usepackage{bm}
\usepackage{array, boldline, makecell, booktabs}

\usepackage{appendix}

\newcommand{\RN}[1]{%
  \textup{\uppercase\expandafter{\romannumeral#1}}%
}

\newcommand{\aven}{\bar{n}}
\newcommand{\bx}{\mathbf{x}}
\newcommand{\br}{\mathbf{r}}

\newcommand{\bq}{\mathbf{q}}

\def\ifmonospace{\ifdim\fontdimen3\font=0pt }

\def\C++{%
\ifmonospace%
    C++%
\else%
    C\kern-.1667em\raise.30ex\hbox{\smaller{++}}%
\fi%
\spacefactor1000 }

\def\Csharp{%
\ifmonospace%
    C\#%
\else%
    C\kern-.1667em\raise.30ex\hbox{\smaller{\#}}%
\fi%
\spacefactor1000 }

\begin{document}
	\preprint{}
	\title{
			Generalizing the structural phase field crystal approach for modeling solid-liquid-vapor phase transformations in pure materials
	      }
	\author{Daniel L. Coelho}
	\email{daniel.coelho@mail.mcgill.ca}
    \affiliation{Department of Physics, Centre for the Physics of Materials, McGill University, Montreal, QC, Canada H3A 2T8}
    \author{Duncan Burns}
    \affiliation{McCormick School of Engineering, Northwestern University, Evanston, IL, United States 60208}
    \author{Emily Wilson}
    \affiliation{Department of Physics, Centre for the Physics of Materials, McGill University, Montreal, QC, Canada H3A 2T8}
	\author{Nikolas Provatas}
	\affiliation{Department of Physics, Centre for the Physics of Materials, McGill University, Montreal, QC, Canada H3A 2T8}
	\date{\today}
	\begin{abstract}

    In a recent class of phase field crystal (PFC) models, the density order parameter is coupled to powers of its mean field. This effectively introduces a phenomenology of higher-order direct correlation functions acting on long wavelengths, which is required for modelling solid-liquid-vapor systems. 
    The present work generalizes these models by incorporating, into a single-field theory, higher-order direct correlations, systematically constructed in reciprocal space to operate across long {\it and} short wavelengths. The correlation kernels introduced are also readily adaptable to describe distinct crystal structures. We examine the three-phase equilibrium properties and phase diagrams of the proposed model, and reproduce parts of the aluminum phase diagram as an example of its versatile parametrization. We assess the dynamics of the model, showing that it allows robust control of the  interface energy between the vapor and condensed phases (liquid and solid). We also examine the dynamics of solid-vapor interfaces over a wide range of parameters and find that dynamical artifacts reported in previous  PFC models do not occur in the present formalism. Additionally, we demonstrate the capacity of the proposed formalism for computing complex microstructures and defects such as dislocations, grain boundaries, and voids in solid-liquid-vapor systems, all of which are expected to be crucial for investigating rapid solidification processes.
	 
	\end{abstract}
	\maketitle

    \section{Introduction}\label{sec:intro}

    Rapid solidification is a key process at work during casting and additive manufacturing processes of metals and alloys, as it plays an important role in shaping the spatio-temporal evolution of material microstructures~\cite{SEKHAR202010}. The interaction between these microstructures, such as grain boundaries, defects, density heterogeneity, and voids~\cite{bourell2016perspectives}, determine the as-cast morphology and can be controlled for achieving specific material performance. Accurately describing these interactions requires the ability to simulate phase transformations from atomic to meso-length scales, over large density and temperature ranges, and among the three phases: solid, liquid, and vapor. These are necessary when investigating the formation of microscopic voids and cracks initiated at dislocations and grain boundaries in solids, which have been shown to substantially impact mechanical properties (strength, ductility, etc.), therefore dictating material performance~\cite{ganchenkova2014effects,fotopoulos2023thermodynamic}. 

The {\it phase field crystal} (PFC) methodology has been successful at modeling phase transformations on diffusive time scales while simultaneously bridging the aforementioned length scales between the atomic- and meso-scales in microstructure modeling. At the heart of the PFC approach is a free-energy functional that can be  minimized by a periodic order parameter in solids, while becoming uniform in disordered phases as in traditional phase field models. This emerges from atomic-length interactions designed into the PFC free-energy density. These can be made to emulate the short wavelength interactions that are motivated by the direct correlation function of classical Density Functional Theory (DFT)~\cite{Ramakrishnan1979,Nik_book}. This simple change of representing a solidifying system in terms of a periodic order parameter allows for a straightforward description of complex microstructures, including polycrystalline grain growth, grain boundaries, anisotropic interface structure, elastic and plastic deformation~\cite{Elder2004}.

The original PFC model~\cite{elder2002,elder2007} free energy consists of ideal and excess free energy contributions. The latter comes from the inclusion of a two-point direct correlation expanded to fourth order in gradients of the order parameter, in such a combination that favors the formation of period phases with single length scale at increasing average density (or low enough effective temperature). The basic PFC formalism was later expanded to allow for a robust control of a variety of two dimensions (2D) and three dimensions (3D) crystal structures, making it possible to model crystallization into and between a wide variety of solids phases of pure metals~\cite{greenwood2010} and the alloys~\cite{Nana2011}. This second PFC approach was coined the {\it structural phase field crystal} (XPFC).

Since the introduction of the original PFC models, several works have further extended the PFC/XPFC formalism to include two uniform phases, liquid {\it and} vapor, an important feature for modelling microscopic voids in the study of pure materials and alloys \cite{kocher2015,Nan2017, Wang2018,wang2020minimal}. Beyond expanding the physical phase space accessible to PFC modeling, this innovation also enabled the coexistence of a solid and a low-density (vapor) phase at temperatures below the triple point, opening the door to the study of  phenomena involving voiding \cite{jreidini2021}, cavitation \cite{jreidini2022}, density shrinkage and hot cracking \cite{Nan2017} and solidification through vapor deposition. These processes are of particular importance in rapid solidification, where the interplay of interface kinetics, defect flow and elasticity is dominant. To date, PFC  modeling of solid-liquid-vapor systems has been limited to crystal lattices with a single length scale in reciprocal space, which can be represented as a real-space fourth order gradient expansion. It is instructive for setting the context of the present work to review below some challenges of the aforementioned vapor-forming PFC models. 


A model by Kocher \textit{et al.}~\cite{kocher2015} featured couplings of the PFC density field with powers of its mean field, effectively introducing higher order correlations operating at long wavelengths (low--$q$; where $q$ is wavenumber in reciprocal space) through a smoothing parameter. This modeling approach was extended to use XPFC kernels, which allowed for efficient control over metallic crystal structures by Jreidini \textit{et al.}~\cite{jreidini2021,jreidini2022}. While expanding the scope of material systems amenable to PFC modeling, the aforementioned models also possessed certain limitations. Namely: (a) they produced solid-liquid-vapor phase diagrams over a limited range of density space, (b) they allowed limited control of the solid-liquid density jump, (c) they exhibited stable interface dynamics only if an upper bound was imposed to the range of the smoothing; smoothing over larger length scales lead to artificial beading at the solid-vapor interfaces. Another approach for modeling the vapor phase was introduced in~\cite{Wang2018}, which used various high-order real space gradients to describe the excess energy. This approach fixed problem (b) but only at the cost of exacerbating problem (a). Moreover, the approach for determining the number and power of the high-order gradients cannot be easily generalized. A more recent approach was introduced in~\cite{frick2023consistent}, which proposed a two-field solution for separately representing spatial variations in the average density and atomic scale density variations in the solid. While this approach was useful in resolving all the aforementioned challenges, it is a hybrid between PFC and traditional phase field theory, rather than a self-consistent single-field PFC theory; this approach is analogous to Schwalbach \textit{et al.} who proposed introducing an additional non-conserved order parameter for representing the vapor phase in the PFC formalism, alongside the spatially varying atomic density field~\cite{schwalbach2013}. 

In this work, we introduce a new PFC methodology for modelling phase transformations in solid-liquid-vapor systems through the use of higher-order direct correlations systematically
constructed in reciprocal space to contain {\it both} short and long wavelength contributions, features missing in previous PFC models. We demonstrate that this approach recovers previous vapor-forming PFC models in specific limits. We show that the aforementioned artifacts of previous vapor-forming PFC models are eliminated in the present formalism. As a case study, we illustrate how to select some parameters of the proposed formalism to quantitatively match temperature-density and pressure-temperature phase diagram of pure aluminum (Al), for the liquid-vapor part of the phase diagrams; quantitative control of the solid phase properties will be left for a separate paper. We also examine the stability and parameterize the energy of liquid-vapor and solid-vapor interfaces.  Finally, we present dynamical simulations of solid-liquid-vapor transitions to demonstrate that the new model lends itself to the investigation of defects and void formation in rapid solidification processes. 

The results in the present paper can be straightforwardly extended to 3D, as shown in a recently developed open-source framework for high performance PFC simulations in 3D~\cite{tatu2024} using the recent three-phase PFC model of Refs.~\cite{jreidini2021,jreidini2022}, which inspired the extensions proposed in this work. However, since the goal here is to present the underlying properties of this newly proposed PFC variant, we limit our simulations to 2D. It is also noted that since the correlation functions for this new model are designed in reciprocal space, it follows on the \textit{XPFC} of Greenwood \textit{et al.}~\cite{greenwood2010,greenwood2011}. Such two-point correlation kernels are easily adapted to allow for robust control of a wide range of metallic crystal structures, ranging from triangular and square lattices, in 2D, to body-centered cubic (BCC), face-centered cubic (FCC), and hexagonal close packed (HCP) crystal lattice structures, in 3D.

This remainder of this paper is organized as follows: 
Section.~\ref{sec:model} introduces the formalism of two-point \textit{and} higher-order direct correlations functions used in the proposed model. Sec.~\ref{sec:vxpfceq} studies the equilibrium properties of the model and presents the construction of temperature-density ($T$-$\bar\rho$) and pressure-temperature ($P$-$T$) phase diagrams for solid-liquid-vapor coexistence over experimentally relevant ranges of temperature and density.  Section~\ref{sec:vxpfcdyn} addresses dynamical artifacts at solid-vapor interfaces in previous models, and examines the stability and energy of interfaces in the proposed model. This section also presents dynamical simulations of the model, showcasing microstructure evolution in several types of solid-liquid-vapor phase transitions. A summary and final remarks of this work are presented in Sec.~\ref{sec:conclusion}.

    \section{Model}\label{sec:model}

    In this section, we present the proposed vapor-forming structural phase field crystal model, which for ease of notation we will hereafter refer to as {\it VXPFC}. Quantities entering the formulation of VXPFC theory are summarized in Table~\ref{tab:pfcquantities}. In accordance with previous approaches that follow the approach of classical density functional theory (CDFT)~\cite{Ramakrishnan1979,evans1979nature,emmerich2012phase}, the starting point is a free-energy functional written in terms of a scaled local density difference denoted by $n(\br)=(\rho(\br)-\rho_0)/\rho_0$, where $\rho(\br)$ is the number density. The free-energy functional accounts for ideal and excess free-energy contributions, and is defined as
\begin{align}
    \Delta\mathcal{F}[n(\br)]\equiv
    \Delta\mathcal{F}_\text{ideal}
    +\sum_{m\,=\,2}^{4}\Delta\mathcal{F}_\text{excess}^{(m)} \,,
    \label{eq_gen_vxpfc}
\end{align}
where $\Delta\mathcal{F}\equiv \mathcal{F}-\mathcal{F}_0\equiv\Delta{F}/{(k_B T_0\rho_0 a^d)}$ is the dimensionless Helmholtz free-energy difference, where $\Delta F = F-F_0$, with $F_0 \equiv f_0 \, V$, where $f_0$ is the free-energy density at the reference point of the functional expansion, defined by the reference temperature $T_0$ and reference density $\rho_0$, and $V$ is the volume of the system. Here, the spatial coordinate is rescaled by the reference solid-phase lattice constant as $\br = \bx/a$, and $d$ represents the dimension of space. The space coordinates are represented as $\bx=(x,y,z)$ for the case where $d=3$, for example.

The ideal gas (noninteracting) free-energy term is typically expanded up to fourth order around the reference density~\cite{elder2002,elder2007}, $n(\br)= 0$, in the spirit of Landau theory, yielding a tractable model for dynamical simulations over relevant density ranges, defined as
\begin{align}
    \Delta\mathcal{F}_\text{ideal}&\equiv \,\tau\int{d\br}[(n(\br)+1)\ln(n(\br)+1)-n(\br)]
    \nonumber \\
    &\approx\,\tau\int{d\br}\big[\tfrac{1}{2}n(\br)^2-\tfrac{1}{6}n(\br)^3+\tfrac{1}{12}n(\br)^4\big] \,,
    \label{ideal_free_energy_term}
\end{align}
where $\tau=T/T_0$ represents the dimensionless temperature.  

Continuing to follow the approach of CDFT, the excess free energy is encapsulated via contributions that employ two-point and higher-order direct density correlation functions, defined as
\begin{align}
    \label{eq:highercorrel}
    \Delta\mathcal{F}_\text{excess}^{(m)} \equiv
    -\dfrac{\tau}{m!}\int{d\br_1}n(\br_1)
    \prod_{j\,=\,2}^m\int{d\br_j}(\rho_0 a^d)^{m-1}\\
    \nonumber
    \times\,C^{(m)}(\br_1,\br_2,\dots,\br_m)
    n(\br_2)n(\br_3)\dots n(\br_m) \,,
\end{align}
and the direct correlation functions $C^{(m)}$ are assumed to follow the form 
\begin{align}
    (\rho_0 a^d)^{m-1}
    C^{(m)}(\br_1,\br_2,\dots,\br_m)\mspace{160mu}
    \nonumber\\
    \equiv Q_m
    \sum_{\{\xi\}}
    C_m(\xi_{1})C_m(\xi_{2})C_m(\xi_{k})\dots C_m(\xi_{m-1}) \,,
    \label{eq:realcorrel}
\end{align}
where the sum can generally run over all permutations of pairwise distances (without repetitions) $\xi_i = \br_{\alpha\beta} = \br_{\alpha} - \br_{\beta}$ ($\alpha, \beta \in \{1,2,3, ...,m\}$), where $i=1,2,3, \cdots m-1$. The $C_m$ can in theory all be different, but to keep things tractable in the construction of a VXPFC model, we consider interactions where they are all the same. Equation~\eqref{eq:realcorrel} represents one of various non-unique methodologies of representing higher-order correlation functions as products of two-point correlations, and which have been shown to be useful in PFC modeling of complex crystal symmetries in recent years \cite{seymour2016,alster2017,wang2018angle}. All the pairwise particle interactions aforementioned can be interpreted as ``indirect'' correlations propagated via increasingly large numbers of intermediate particles \cite{hansen1990theory}, Chapter 3.
The base two-point correlation kernels $C_m$, from which $C^{(m)}$ in Eq.~(\ref{eq:realcorrel}) is constructed, are connected to their reciprocal space counterparts via
\begin{align}
    C_{m}(|\br_i- \br_j|)
    =
    \int 
    \dfrac{d\bq}{(2\pi)^{d}} \hat{C}_{m}(\bq)
    e^{-i\bq \cdot (\br_i-\br_j)} \,,
    \label{eq:fouriercorrel}
\end{align}
where $\bq$ is a general wave vector, and $\hat{C}_{m}(\bq)$ is the Fourier transform of $C_m(|\br_i- \br_j|)$. We note that there are several ways to express the higher order direct correlation kernels that are consistent with the notation of Eq.~(\ref{eq:realcorrel}). 
\begin{table}[t]
    \caption{\label{tab:pfcquantities} Physical quantities used in the dimensionless parameters of the proposed VXPFC theory. Some quantities are chosen to match the properties of pure aluminum (Al), and the reference density is taken to be that of  liquid Al at the triple point, $\tilde{\rho}_0=2,368\,\text{kg}\cdot\text{m}^{-3}$.}
    \begin{ruledtabular}
    \begin{tabular}{lll}
      \textbf{Quantity} & \textbf{Symbol} & \textbf{Units} \\[2pt]
      \hline \\[-8pt]
      Helmholtz free energy & $F$ & J \\
      Boltzmann constant & $k_B$ & $\text{J}{\cdot}\text{K}^{-1}$ \\
      Avogadro's number & $N_A$ & $\text{mol}^{-1}$\\
      Gas constant & $R=N_Ak_B$ & $\text{J}{\cdot}\text{mol}^{-1}{\cdot}\text{K}^{-1}$ \\
      Temperature & $T$ & $\text{K}$ \\
      Reference temperature & $T_0$ & $\text{K}$ \\
      Reduced temperature & $\tau=T/T_0$ & - \\
      Reference particle density & $\rho_0=\tilde{\rho}_0(N_A/m_i)a^{3-d}$ & $\text{m}^{-d}$ \\
      Reference mass density & $\tilde{\rho}_0$ & $\text{kg}{\cdot}\text{m}^{-3}$ \\
      Lattice spacing & $a$ & \text{m} \\
      Dimensionless position & $\br=\bx/a$ & -\\
      Dimension of space & $d\;(=1,2,3)$ & - \\
      Molar mass for $i$ species & $m_i$ & $\text{kg}{\cdot}\text{mol}^{-1}$ \\
      $m$-point direct & 
      \multirow{ 2}{*}{
      $C^{(m)}(\br_1,\br_2,\dots,\br_m)$} & 
      \multirow{ 2}{*}{-}
      \\
        correlation function & & 
    \end{tabular}
    \end{ruledtabular}
\end{table}

In the general PFC framework, the contribution due to the two-point correlation ($m=2$) is the lowest order term used for establishing a periodic lattice structure of the ordered (solid) phase. Additionally, its low-$q$ properties also allow for minimal control of the compressibility of the liquid.  One can also modify the coefficients of some local polynomial terms arising from the ideal free energy to allow for a second uniform phase (vapor). This can be formally justified by incorporating long-wavelength contributions in the two-point ($m=2$), three-point ($m=3$) and four-point ($m=4$) correlation functions added to the excess free energy, each including a corresponding self-interaction term (i.e., a local polynomial term). In order to allow for this, we  hereafter define
\begin{align}
    \hat{C}_{m}(\bq)=
    \hat{C}_{m}(0)+
    \hat{\tilde{C}}_{m}(\bq) \,,
    \label{eq:kernelexp}
\end{align}
where $\hat{C}_{m}(0)$ corresponds to the value of the correlation kernel at $\bq=0$, conveniently introduced to yield the self-interaction terms when incorporated into the respective excess free-energy terms. Powers of $\hat{C}_{m}(0)$ emerging from the excess free-energy terms will re-scale the coefficients of the ideal free energy at the corresponding orders ($m=2,3,4$). The $\bq\neq 0$  reciprocal space dependence is encoded in $\hat{\tilde{C}}_{m}(\bq)$, which in real space represents the spatial dependence of the correlation functions (this will be demonstrated in the following sections). The inclusion of high-$q$ contributions to the higher order correlation functions will prove to be useful for providing robust control of the solid phase in relation to the uniform phases. This is analogous to the recent work of Huang and co-workers~\cite{wang2018angle,wang2020minimal,wang2022control}, albeit using a different approach. The next three subsections elucidate the specific forms of excess energy terms generated by the $m=2, 3, 4$ correlations functions.

We note that the formalism of constructing higher order correlations as products of two-point correlation kernels has been studied in a limited fashion in previous PFC works. Seymour \textit{et al.}~\cite{seymour2016} proposed the first rotationally invariant three-point correlation function (3PCF) comprising two terms of the form, $C^{(3)}(\br_{1},\br_{2},\br_{3})\propto C_3(\br_{12})C_3(\br_{13})$, whose short wavelength properties stabilize a wide variety of anisotropic crystalline structures, including graphene and Kagome lattices (as well as square and triangular lattices, which can also be controllable solely through the 2PCF in the usual XPFC approach~\cite{greenwood2010,greenwood2011}). Jreidini \textit{et al.} later showed how this approach produces rectangular, rhombic lattices in 2D \cite{Paulthesis}. Alster \textit{et al.}~\cite{alster2017} and Seymour~\cite{Seymourthesis} extended this 3PCF approach to 3D to capture simple cubic, diamond cubic, graphene layers and CaF$_2$ lattices \cite{Alster20173PCF}. This approach of making higher order correlation functions as products of two-point correlation functions is also inherent in the design of the original vapor PFC models by Kocher \textit{et al.}~\cite{kocher2015} and Jreidini \textit{et al.}~\cite{jreidini2021,jreidini2022}, except  that the $C_3(\br_{\alpha\beta})$ of these were designed to capture \textit{only} long-wavelength correlations of the density field. As mentioned above, the present work generalizes this approach to generate two-, three-, and four-point correlation functions operating at both long and short wavelengths.  Toward this aim, we design a class of $\hat{C}_m(\bq)$ that allows for both the coexistence of and dynamical transitions between solid-liquid-vapor phases.

\subsection{Two-point direct correlation ($m=2$)}

\begin{figure}[t]
	\includegraphics[width=.4\textwidth]{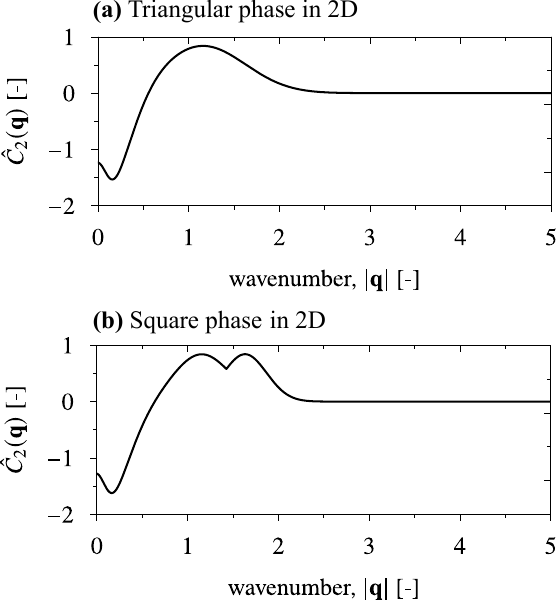}
	\caption{\label{fig:2PCF} Two-point direct correlation functions used in this work represented by Eq.~(\ref{basic_XPFC_C2}) with the parameters  indicated in Table~\ref{tab:modelparams} for building (a) triangular, and (b) square solid phases, in two dimensions (2D).}
\end{figure}
From Eq.~(\ref{eq:realcorrel}), the two-point direct correlation function (2PCF) term becomes $(\rho_0 a^d)C^{(2)}(\br_1,\br_2)\equiv C_2(|\br_1-\br_2|)$, where $Q_2$ has been set to one here. Substituting this definition into Eq.~\eqref{eq:highercorrel} gives,
\begin{align}
  \Delta\mathcal{F}_\text{excess}^{(2)}=&
    -\dfrac{\tau}{2}\int{d\br_1}\,n(\br_1)
    \int{d\br_2}
     C_2(|\br_1-\br_2|)
    n(\br_2) \,,
    \label{2_point_excess_term}
\end{align}
which is the starting form from which the excess energy is specialized in all previous PFC models. Originally, $\hat{C}_2(\bq)$ was designed as an polynomial expansion~\cite{Elder2004,elder2007} truncated to fourth order in $\bq$ to place the required positive peak at $|\bq_0|$ to stabilize a periodic crystal structure of interest. Additionally, a peak at $\bq=0$ was added to model liquid state compressibility~\cite{elder2002,elder2007,Nik_book}. The two-point correlation has also been truncated at higher order~\cite{wang2018angle,wang2020minimal,wang2022control}, producing a peak at $\bq=0$ with {\it negative} curvature $\hat{C}_2^{''}(\bq)$, which is required to stabilize vapor-liquid interfaces \cite{omelyan2004}. The original XPFC approach \cite{greenwood2010,greenwood2011} modelled the peaks of $\hat{C}_2(\bq)$ with one or more Gaussian functions situated at the peaks ($|\bq_0|, |\bq_1|, |\bq_2|, \cdots$) corresponding to the primary lattice reflections of the crystal structure of interest. The position, height, and curvature of each peak can be tuned independently. 
Here, we extend the XPFC form of $\hat{C}_2(\bq)$ by adding {\it two} Gaussian peaks at $\bq=0$, such as to produce a negative curvature $\hat{C}_2^{''}(\bq=0)$, while keeping $\hat{C}_2(\bq=0)<0$. These double Gaussian functions are designed to (a) control the absolute value of the correlation at $\bq=0$, and in turn the compressibility of the uniform phases, and (b) produce a negative curvature around $\bq=0$ to yield a $|\nabla\aven|^2$ type behavior, where $\bar{n}$ is the locally averaged density field. Point (b) is useful to describe energy contributions due to average density gradients across solid-vapor, liquid-vapor, and solid-liquid interfaces more rigorously~\cite{omelyan2004microscopic}. With above considerations in mind, we introduce the following two-point direct correlation function, defined in Fourier space, 
{
\begin{align}
\hat{C}_2(\bq) = &\;\,
\kappa_{1,2}
e^{-\frac{|\bq|^2}{2\beta^2}}
-\kappa_{2,2}
e^{-\frac{|\bq|^4}{2\gamma^2}}
\nonumber\\&
+\text{max}_i\bigg( B_{2,i}^x
e^{-\frac{|\bq_i|^2}{2\sigma^2}\tau}\, 
e^{-\frac{(|\bq|^2-|\bq_{i}|^2)^2}{2\alpha_{i}^2}}\bigg) \,,
\label{basic_XPFC_C2}
\end{align}
where the $\text{max}_i$ implies the maximum envelope over all terms containing the index $i$.} For the case of 2D examined here,  $i=\{10\}$ and $\{11\}$, corresponding to peaks at $|\bq_{10}|$ and $|\bq_{11}|$, and the prefactor $\exp[-(|\bq_i|^2/2\sigma_i^2)\tau]$ carries both a Debye-Waller-like factor (temperature dependent) and an elastic compressibility constant. All parameters in  Eq.~\eqref{basic_XPFC_C2} and their value used in this work are shown in Table~\ref{tab:modelparams}. Following the definition in Eq.~\eqref{eq:kernelexp}, we write $\hat{C}_{2}(\bq)=\hat{C}_{2}(0)+\hat{\tilde{C}}_{2}(\bq)$, and substitute this form into Eq.\eqref{2_point_excess_term} via Eq.~\eqref{eq:fouriercorrel}. This yields the following excess free energy contribution from the two-point correlation,
\begin{align}
  \Delta\mathcal{F}_\text{excess}^{(2)}=&
    -\dfrac{\tau}{2}\int{d\br_1}
    \left[
    \hat{C}_2(0)n^2(\br_1) + 
    n(\br_1)\eta_{(2)}(\br_1)
    \right] \,,
    \label{2_point_excess_term_new}
\end{align}
where 
\begin{align}
\eta_{(2)}(\br) = \{\tilde{C}_2 \ast n\}(\br) &\equiv \int d\br' \tilde{C}_2(\br-\br')n(\br')\nonumber \\
&=[\hat{\tilde{C}}_{2}(\bq)\hat{n}(\bq)]_{\br} \,.
\end{align}
{ where the inverse Fourier transform is denoted by $[\cdot]_{\br}$.} We highlight once again the contribution due to a self-interaction (quadratic) term and the $\eta_{(2)}(\br)$ term operating at {\it both} long and short wavelengths. The former term corresponds to that of $\hat{C}_{2}$ at $\bq=0$, which will re-scale the quadratic term in the ideal free energy in the final form of the model, summarized in Sec.~\ref{sec:vxpfc}. Figure~\ref{fig:2PCF} illustrates two forms of the two-point correlation function in Eq.~(\ref{basic_XPFC_C2}) used in this work, using the parameters indicated in Table~\ref{tab:modelparams}.

\subsection{Three-point direct correlation ($m=3$)}

Peebles and Groth have proposed a generalized hierarchical form of multipoint correlation functions comprising products of lower-order correlation functions to describe structure in cosmological fields~\cite{peebles1975statistical,groth1977statistical}. The specific form they proposed for the 3PCF is $C^{(3)}(\br_{1},\br_{2},\br_{3})\propto C_3(\br_{12})C_3(\br_{31})+C_3(\br_{12})C_3(\br_{23})+C_3(\br_{23})C_3(\br_{31})$, which allows for different contributions of $\mathcal{O}(n^3)$ to be incorporated in the model. 

\begin{figure}[t]
	\includegraphics[width=.45\textwidth]{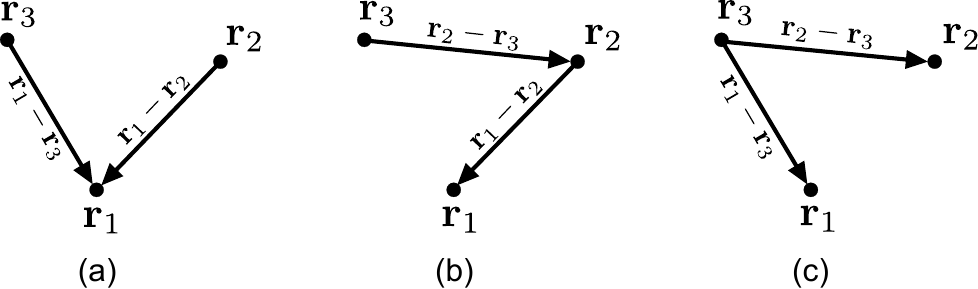}
	\caption{\label{fig:highercorrel} All possible pairwise distances considered in the representation of the three-point correlation, considering translational and rotational invariance. (a) $C_3(\br_1-\br_2)C_3(\br_1-\br_3)$, (b) $C_3(\br_1-\br_2)C_3(\br_2-\br_3)$, and (b) $C_3(\br_1-\br_3)C_3(\br_2-\br_3)$.}
 \label{pairwise_3point_fig}
\end{figure}

To model the 3PCF, we adopt the phenomenology of~\cite{peebles1975statistical,groth1977statistical} and, following the definition in Eq.~\eqref{eq:realcorrel}, propose the form
\begin{align}
\label{3_point_cor}
    (\rho_0 &a^d)^{2}
    C^{(3)}(\br_1,\br_2,\br_3)
    \nonumber\\
    =&\;Q_3\big[C_3(\br_1-\br_2)C_3(\br_1-\br_3)
    +C_3(\br_1-\br_2)C_3(\br_2-\br_3)
    \nonumber\\&
    +C_3(\br_1-\br_3)C_3(\br_2-\br_3)\big]\,,
\end{align}
where $\{\xi\}=\{\br_{12},\br_{13},\br_{23}\}$, and as alluded to above, it is tacitly understood that $C(\br_i-\br_j)\equiv C(|\br_i-\br_j|)$ throughout. Substituting this definition into the general form of Eq.~\eqref{eq:highercorrel}, yields an excess free energy term of the form, 
\begin{align}
\label{eq:3pointexcess}
    \Delta\mathcal{F}_\text{excess}^{(3)}=&
    -\dfrac{\tau}{3!}\int{d\br_1}n(\br_1)
    \iint d\br_2 d\br_3(\rho_0 a^d)^{2}
    \nonumber\\&
    \times C^{(3)}(\br_1,\br_2,\br_3)
    n(\br_2)n(\br_3) \,
    \nonumber\\
    =&-\dfrac{\tau}{3!}Q_3\int{d\br_1}n(\br_1)
    \iint d\br_2 d\br_3 \big[
    \nonumber\\&
    C_3(\br_1-\br_2)C_3(\br_1-\br_3)
    +C_3(\br_1-\br_2)C_3(\br_2-\br_3)
    \nonumber\\&
    +C_3(\br_1-\br_3)C_3(\br_2-\br_3)\big]
    n(\br_2)n(\br_3)\,.
\end{align}
An illustration of what the three correlation products represent in terms of statistical interactions between points in the density field is shown in Fig.~\ref{fig:highercorrel}. To proceed further, we follow the definition in Eq.~\eqref{eq:kernelexp} once again and write the base correlation kernel $C_3(\xi_i)$ in Fourier space as $\hat{C}_{3}(\bq)=\hat{C}_{3}(0)+\hat{\tilde{C}}_{3}(\bq)$. For simplicity, we take $\hat{C}_{3}(\bq)$ to have the form in Eq.~(\ref{basic_XPFC_C2}) (it is generalized in Eq.~(\ref{basic_XPFC_Cm}) to order $m$), with coefficients given in Table~\ref{tab:modelparams}. Considering first the term $C_3(\br_1-\br_2)C_3(\br_1-\br_3)$ in Eq.~\eqref{eq:3pointexcess}, this decomposition gives the contribution

\begin{align}
    \Delta\mathcal{F}_{\text{excess}}^{(3)}=
    &-\dfrac{\tau}{3!}Q_3\int{d\br}
    \big[
    \hat{C}_3^2(0) n^3(\br)
    \nonumber\\
    &+2\hat{C}_3(0) n^2(\br)\eta_{(3)}(\br_1)
    + n(\br)\eta_{(3)}^2(\br)
    \big] \,,
    \label{3_point_1st_term}
\end{align}
where
\begin{align}
\eta_{(3)}(\br) = \{\tilde{C}_3 \ast n\}(\br) &\equiv \int d\br' \tilde{C}_3(\br-\br')n(\br')\nonumber \\
&=[\hat{\tilde{C}}_{3}(\bq)\hat{n}(\bq)]_{\br} \,, 
\end{align}
and where $Q_3$ also incorporates a factor of three due to the multiplicity of the other interaction terms, $C_3(\br_1-\br_2)C_3(\br_2-\br_3)$ and $ C_3(\br_1-\br_3)C_3(\br_2-\br_3)$, each producing an identical contribution to the free energy as Eq.~\eqref{3_point_1st_term}, as is also evident from the symmetry of Fig.~\ref{pairwise_3point_fig}. It thus suffices to retain only  Eq.~\eqref{3_point_1st_term} for the 3PCF contribution to the excess energy. The values used for these parameters are shown in Table~\ref{tab:modelparams}. 


\subsection{Four-point direct correlation ($m=4$)}
Continuing analogously to the previous subsection, and, following the definition in Eq.~\eqref{eq:realcorrel}, the four-point correlation function (4PCF) is defined by
\begin{align}
\label{4_point_cor}
    (\rho_0 &a^d)^{3}
    C^{(4)}(\br_1,\br_2,\br_3,\br_4)
    \nonumber\\
    =&\;
    Q_4\big[C_4(\br_1-\br_2)C_4(\br_1-\br_3)C_4(\br_1-\br_4)
    \nonumber\\
    &+C_4(\br_1-\br_2)C_4(\br_1-\br_3)C_4(\br_2-\br_4)+\dots \big]\,,
\end{align}
where $\{\xi\}=\{\br_{12},\br_{13},\br_{14},\br_{23},\br_{24},\br_{34}\}$, and where the ($\dots$) represent all the triple correlation products that can be formed from the set $\{\xi\}$. Using this definition, the general form of the excess free energy due to the three-point correlation becomes
\begin{align}
    \Delta\mathcal{F}_\text{excess}^{(4)}=&
    -\dfrac{\tau}{4!}Q_4\int{d\br_1}n(\br_1)
    \int d\br_2 d\br_3d\br_4(\rho_0 a^d)^{3}
    \nonumber\\&
    \times C^{(4)}(\br_1,\br_2,\br_3,\br_4)
    n(\br_2)n(\br_3)n(\br_4) \,. 
    \nonumber\\
    =&-\dfrac{\tau}{4!}Q_4\int{d\br_1}n(\br_1)
    \int d\br_2 d\br_3d\br_4 
    \nonumber\\&
    \times \big[
    C_4(\br_1-\br_2)C_4(\br_1-\br_3)C_4(\br_1-\br_4) +\dots
    \nonumber\\&
    +C_4(\br_1-\br_2)C_4(\br_2-\br_3)C_4(\br_3-\br_4) +\dots
    \nonumber\\&
    +C_4(\br_1-\br_2)C_4(\br_1-\br_3)C_4(\br_2-\br_4)+\dots
    \nonumber\\&
    +C_4(\br_1-\br_2)C_4(\br_2-\br_4)C_4(\br_3-\br_4) +\dots\big]
    \nonumber\\&
    \times n(\br_2)n(\br_3)n(\br_4) \,,
    \label{excess_4point_term}
\end{align}
where $(\dots)$ in each line denotes $4$ cyclic permutations on the indices of the indicated term, each yielding an identical contribution to the excess free energy. { It is noted that we do not include terms with products missing one of the vector positions ($\br_1$, $\br_2$, $\br_3$, or $\br_4$), such as the ones of the type $C(\br_{12}) C(\br_{13}) C(\br_{23})$, as these do not link all four vector positions.} Similarly to the 2PCF and 3PFC excess terms, we decompose the base correlation kernel by following the definition in Eq.~\eqref{eq:kernelexp}, and write $C_4(\xi_i)$ in Fourier space as $\hat{C}_{4}(\bq)=\hat{C}_{4}(0)+\hat{\tilde{C}}_{4}(\bq)$. As previously, we take the full correlation kernel $\hat{C}_{4}(\bq)$ to have the form in Eq.~(\ref{basic_XPFC_C2}) (or, alternatively, Eq.~(\ref{basic_XPFC_Cm}), with $m=4$), with the coefficients given in Table~\ref{tab:modelparams}. Substituting this into the  $C_4(\br_1-\br_2)C_4(\br_1-\br_3)C_4(\br_1-\br_4)$ term of Eq.~(\ref{excess_4point_term}) yields, after some manipulations,
\begin{align}
    \Delta\mathcal{F}_\text{excess}^{(4)}
    =&-\dfrac{\tau}{4!}Q_{4}\int{d\br}
    \bigg[\hat{C}_4^3(0) n^4(\br)
    \nonumber\\&
    +3\hat{C}_4^2(0)n^3(\br)\eta_{(4)}(\br)
    \nonumber\\&
    +3\hat{C}_4(0)n^2(\br)\eta_{(4)}^2(\br)
    +n(\br)\eta_{(4)}^3(\br)\bigg] \,,
    \label{excess_4_term}
\end{align}
where 
\begin{align}
    \eta_{(4)}(\br) = \{\tilde{C}_4 \ast n\}(\br) &\equiv \int d\br' \tilde{C}_4(\br-\br')n(\br') \nonumber \\
    &=[\hat{\tilde{C}}_{4}(\bq)\hat{n}(\bq)]_{\br} \,,
\end{align}
and where $Q_4$ also incorporates a factor of four due to the multiplicity of the other cyclic permutations of the term $C_4(\br_1-\br_2)C_4(\br_1-\br_3)C_4(\br_1-\br_4)$, which yield the same result as Eq.~(\ref{excess_4_term}) and are thus not required further than in their formal representation in Eq.~(\ref{excess_4point_term}). The values used for these parameters are shown in Table~\ref{tab:modelparams}. Also, in this work, we will not consider contributions from the four-point correlation terms arising from the last three lines of Eq.~(\ref{excess_4point_term}). These represent three different groups of correlation products with distinct pairwise distances which can be shown to contribute terms of the form $n^2\nabla^2n^2$ or $n^2\nabla^4n^2$ in the free energy, which are indicated in Ref.~\cite{wang2018angle} to be necessary for controlling angular dependence of crystal structure.

\subsection{Structural PFC model for solid-liquid-vapor systems (VXPFC)}\label{sec:vxpfc}
In the last sections, we defined each term of the generating CDFT free-energy functional in Eq.~\eqref{eq_gen_vxpfc}. Now, we proceed to summarize the final form of the free energy considered in this work, given by
\begin{align}
    \Delta\mathcal{F}\equiv&\;\,
    \Delta\mathcal{F}_\text{ideal}
    +\Delta\mathcal{F}_\text{excess}^{(2)}
    \nonumber\\&
    +\Delta\mathcal{F}_\text{excess}^{(3)}
    +\Delta\mathcal{F}_\text{excess}^{(4)}
    +\Delta\mathcal{F}_\text{linear}\,,
    \label{eq:vxpfcmodel3}
\end{align}
where we tacitly introduced a linear term with coefficient $B_0$, $\Delta\mathcal{F}_\text{linear}=-\tau\int{d\br} B_0 n(\br)$, interpreted as an external contribution to the free energy, which was recently demonstrated to be essential for the calculation and control of the system's bulk pressure and elastic constants~\cite{wang2020minimal,Burns2023_2}. Collecting Eq.~(\ref{ideal_free_energy_term}) and the higher order excess terms proposed in Eq.~(\ref{2_point_excess_term_new}), Eq.~(\ref{3_point_1st_term}) and Eq.~(\ref{excess_4_term}) into Eq.~(\ref{eq:vxpfcmodel3}) yields the final form of the VXPFC model for solid-liquid-vapor systems. After some tedious but straightforward algebra, this is written in the following compact form, 
\begin{align}
    \Delta\mathcal{F} =&\;
    -\tau\int{d\br}
    \big[B_0 n(\br)\big]
    \nonumber\\&
    -\dfrac{\tau}{2!}\int{d\br}  
    \,\big[
    C_0n^2(\br) + 
    n(\br)\eta_{(2)}(\br)
    \big]
    \nonumber\\&
    -\dfrac{\tau}{3!}\int{d\br}
    \big[
    D_0n^3(\br) +
    D_1 n^2(\br)\eta_{(3)}(\br) 
    \nonumber\\&
    +D_2 n(\br)\eta_{(3)}^2(\br)
    \big]
    \nonumber\\&
    -\dfrac{\tau}{4!}\int{d\br}
    \big[
    E_0 n^4(\br)
    +E_1 n^3(\br)\eta_{(4)}(\br)
    \nonumber\\&
    +E_2 n^2(\br)\eta^2_{(4)}(\br)
    +E_3 n(\br)\eta^3_{(4)}(\br)
    \big] \,,
    \label{eq:vxpfcmodel2}
\end{align}
where 
\begin{align}
    \eta_{(m)}(\br) = \int d\br' \tilde{C}_m(\br-\br')n(\br')
    =\bigg[\hat{\tilde{C}}_{m}(\bq)\hat{n}(\bq)\bigg]_{\br} \,,
    \label{general_eta_form}
\end{align}
and
\begin{equation}
\hat{\tilde{C}}_{m}(\bq)=\hat{C}_{m}(\bq)-\hat{C}_{m}(0) \,,    
\end{equation}
with $\hat{C}_{m}(\bq)$ given by Eq.~\eqref{basic_XPFC_C2} for the two-point correlation kernel ($m=2$), and
{
\begin{align}
\hat{C}_m(\bq) = &\;\,
\kappa_{1,m}
e^{-\frac{|\bq|^2}{2\beta^2}}
-\kappa_{2,m}
e^{-\frac{|\bq|^4}{2\gamma^2}}
\nonumber\\&
+\text{max}_i\bigg(B_{m,i}^x
e^{-\frac{(|\bq|^2-|\bq_{i}|
^2)^2}{2\alpha_{i}^2}}\bigg) \,,
\label{basic_XPFC_Cm}
\end{align}}
for the three-point ($m={3}$), and four-point correlations kernels ($m={4}$). As a first analysis of this VXPFC model, we follow ~\cite{wang2020minimal,jreidini2022} and assume a temperature dependence of $B_0$ and $C_0$, namely
\begin{align}
    \label{eq_B0C0}
    B_0\equiv B_{00}+B_{01}\tau \,, \nonumber \\
    C_0\equiv C_{00}+ C_{01}\tau \,.
\end{align} 
The parameters in Eqs.~\eqref{eq:vxpfcmodel2}, Eq.~(\ref{basic_XPFC_Cm}) and Eq.~(\ref{eq_B0C0}) are shown/defined in Table~\ref{tab:modelparams}.

We note that for simplicity Eq.~\eqref{basic_XPFC_Cm} does not contain a Debye-Waller-like factor as the temperature dependence is kept in the aforementioned two-point kernel only. More general forms will be considered in future applications.  It is noted that the first vapor PFC model by Kocher \textit{et al.}~\cite{kocher2015}, and the extended version of Jreidini \textit{et al.}~\cite{jreidini2021,jreidini2022}, can be interpreted as particular cases of Eq.~\eqref{eq:vxpfcmodel2}. Namely, both of them couple the density field to powers of its mean field, which can be interpreted as higher-order excess terms in which the correlation functions operate  only over long wavelength ranges, as shown by Eq.~\eqref{eq:kochermodel} and Eq.~\eqref{eq:jreidinimodel}, respectively.  It is also reiterated that if the correlation kernel is designed in real-space by an expansion of gradients, we recover the model of Wang \textit{et al.} \cite{wang2020minimal} in Eq.~\eqref{eq:zfmodel} (See  Appendix~\ref{sec:singlefield}). 
\begin{table}[t]\caption{\label{tab:modelparams} Parameter values used in all numerical calculations made with the model defined by Eqs.~(\ref{eq:vxpfcmodel2}), (\ref{general_eta_form}) and (\ref{basic_XPFC_Cm}). The first set corresponds to values of the physical quantities considered in this work. The second set contains general parameters shared by two-, three-, and four-point correlation functions, while their individual parameters are presented in the subsequent sets, respectively. The last set contains the definitions of the relations between model parameters.}
		\begin{ruledtabular}
		\begin{tabular}{@{}llllll@{}} 
			\textbf{Parameter} & \textbf{Value} & 
			\textbf{Units} &
			\textbf{Parameter} & \textbf{Value} 
			& \textbf{Units}\\[2pt]
            \hline\hline \\[-8pt]
			$T_{0}$  & 933 & K 
            & $T_\text{CP}$  & 6250 & K \\
            ${\rho}_0$ & $5.29\times10^8$ & $\text{m}^{-3}$ &
            $\tilde{\rho}_0$ & 2.368 & $\text{kg}{\cdot}\text{m}^{-3}$ \\
            $k_B$ & $1.38\times10^{-23}$ & $\text{J}{\cdot}\text{K}^{-1}$ &
            $N_A$ & $6.022\times10^{23}$ & $\text{mol}^{-1}$
            \\[3pt]\hline\\[-8pt]
            
            $|\bq_{10}|$  & $2/\sqrt 3$ & - & $|\bq_{11}|$ & $|\bq_{10}|\sqrt{2}$ & -\\
            $\alpha_{10}$ & 1.2 & - &  $\alpha_{11}$ & 0.8 & - \\
            $\beta$ & 0.25 & - & $\gamma$& 0.25 & - \\
            $N_a$ & $0.01$ & - & $\sigma$ & 1.7450 & - \\
            $B_{00}$ & 0.1040 & - & $B_{01}$ & -0.1023 & - 
            \\[3pt]\hline\\[-8pt]
            
            $\kappa_{1,2}$ &1.5& - & 
            \multicolumn{3}{l}{$\kappa_{2,2}=
            \kappa_{1,2} - [1+C_0] + \hat{\tilde{C}}_2(0)$}\\
            $C_{00}$ & -1.703 & - & $C_{01}$ & -0.1667 & - \\
            $B_{2,10}^x$ & 1.0528 & - & $B_{2,11}^x$ & $0.9B_{2,10}^x$ & - \\[3pt]\hline\\[-8pt]
            
            $\kappa_{1,3}$ & 0.2 & - & $\kappa_{2,3}$& 2.2 & - \\
            $D_0$ & -8.139 & - & $D_{1}$ & -9.1390 & - \\
            $D_{2}$ & -2.2845 & - & - & - & -
            \\
            $B_{3,10}^x$ & -0.035 & - & $B_{3,11}^x$ & $0.9B_{3,10}^x$ & -
            \\[3pt]\hline\\[-8pt]
            
            $\kappa_{1,4}$ & 0.2126 & - & $\kappa_{2,4}$& 1.8 & - \\
            $E_0$ & -11.74 & - &$E_{1}$ & -18.4074 & - \\
            $E_{2}$ & -11.5960 & - & $E_{3}$ & -2.4350 & - 
            \\
            $B_{4,10}^x$ & -0.035 & - & $B_{4,11}^x$ & $0.9B_{4,10}^x$ & - 
            \\[3pt]\hline\\[-8pt]
            
             \multicolumn{6}{l}{
             $Q_3=0.25(D_0-1)=-2.2848$,\quad 
             $Q_4=0.25(E_0+2)=-2.4350$}\\[1pt]
            \multicolumn{6}{l}{
             $\hat{C}_2(0)=C_0+1\,$}\\[1pt]
             \multicolumn{6}{l}{
             $\hat{C}_3(0)=\sqrt{(D_0-1)/Q_3}\,$,
             $D_1=2Q_{3}\hat{C}_3(0)\,$, 
             $\;D_2=Q_{3}$}\\[3pt]
             \multicolumn{6}{l}{
             $\hat{C}_4(0)=\sqrt[3]{(E_0+2)/Q_4}$,
             $E_1=3Q_{4}\hat{C}_4^2(0)$, 
             $E_2=3Q_{4}\hat{C}_4(0)$, 
             $E_3=Q_{4}$}
		\end{tabular}
		\end{ruledtabular}
	\end{table}
    
    \section{Equilibrium properties}\label{sec:vxpfceq}

    The VXPFC model contains short-wavelength interactions in the two, three and four-point  correlation functions, which allow for robust and quantitative control over the stability and properties of the solid phase relative to the uniform phases, the latter of which emerge from the long-wavelength properties of said correlation functions. This is demonstrated in this section, which examines the equilibrium properties and phase diagram of the model.  The parameters in Eqs.~\eqref{eq:vxpfcmodel2}--\eqref{basic_XPFC_Cm} are chosen to quantitatively reproduce the aluminum phase diagram in ($T,\bar{\rho}$) space, and the triple point ($T=933$K) and critical point ($T_\text{CP}$) of the aluminum phase diagram in ($T, P$) space.  Data for the aluminum phase diagram are obtained from a multiphase equation of state (EoS) model in~\cite{lomonosov2007multi}. Unless otherwise stated, all parameters used are displayed in Table~\ref{tab:modelparams}. 

The starting point to examine the equilibrium properties of the  model is to approximate the PFC density by a mode approximation (ansatz), which captures the periodic structure of the solid lattices being considered. Following the works in~\cite{elder2002,greenwood2011,wang2020minimal}, the density field is expanded as 
\begin{align}
    n(\br)=
    \aven + \sum_{j}
    \big\{
    A_{j} e^{i\mathbf{G}_j\cdot\br} +
    A_{j}^* e^{-i\mathbf{G}_j\cdot\br}
    \big\}\,,
    \label{eq:one-modeapp}
\end{align}
where $\aven$ corresponds to the phase's average density, $A_j$ is the amplitude (or order parameter) controlling density oscillations along the lattice planes represented by the reciprocal lattice (wave) vectors $\mathbf{G}_j$ of the crystal. In this representation, uniform phases correspond to $A_j=0$, while solid phases correspond to at least one of the $A_j>0$, and whose values are determined from equilibrium considerations, described below.

\subsection{Solid-liquid-vapor phase coexistence in two dimensions: triangular phase}
\begin{figure}[t]
	\includegraphics[width=.45\textwidth]{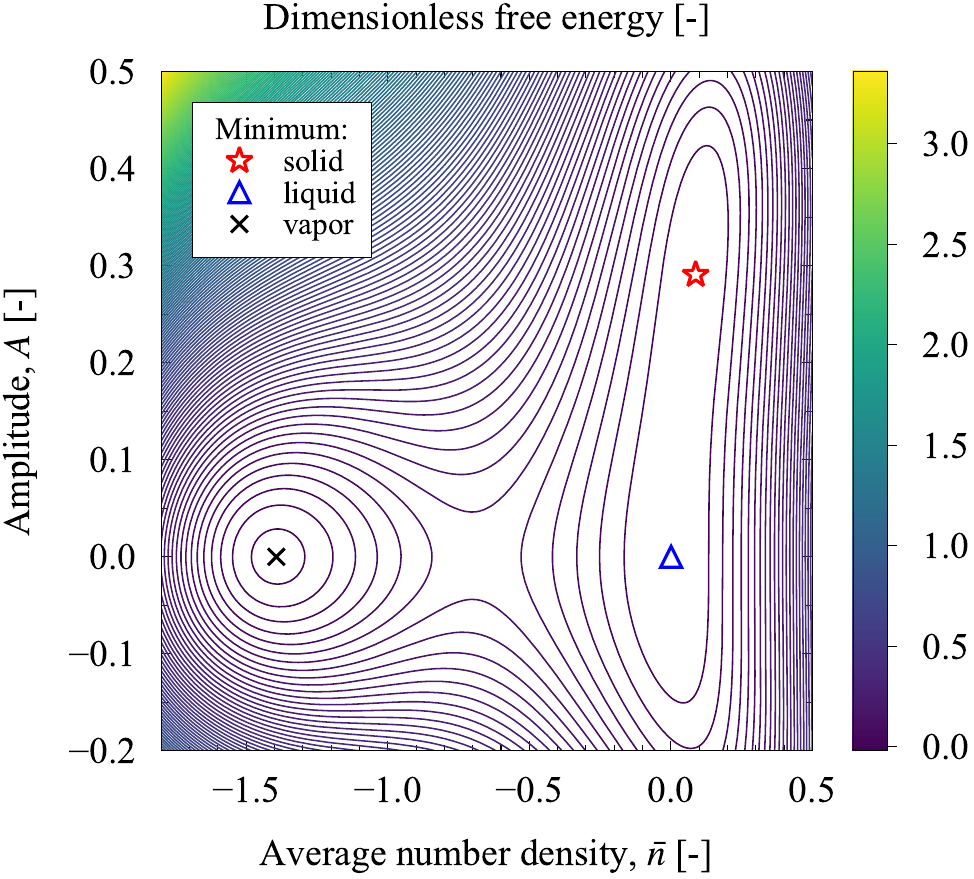}
	\caption{\label{fig:felandscape} The landscape of the VXPFC model free energy in Eq.~\eqref{eq:fetridensity}, after integration over the unit cell of the triangular phase for $\bq=\bq_{10}$, at the triple point $\tau=1.0$ ($T=933$K), using parameters in Table~\ref{tab:modelparams}. The dots are the minimum values corresponding to the vapor (black), liquid (blue) and solid (red) phase wells. The liquid well corresponds to a minimum at $(\aven,A)=(0,0)$ as the reference mass density is that of the liquid phase ($\tilde{\rho}_0=2,368\,\text{kg}\cdot\text{m}^{-3}$), and $\aven=(\bar\rho-\rho_0)/\rho_0$, with a corresponding  reference particle density $\rho_0=\tilde{\rho}_0(N_A/m_i)$.} 
 \label{free_energy_landscape_fig}
\end{figure}

\begin{figure}[tbh]
    \begin{tabular}{c}
         \includegraphics[width=.45\textwidth]{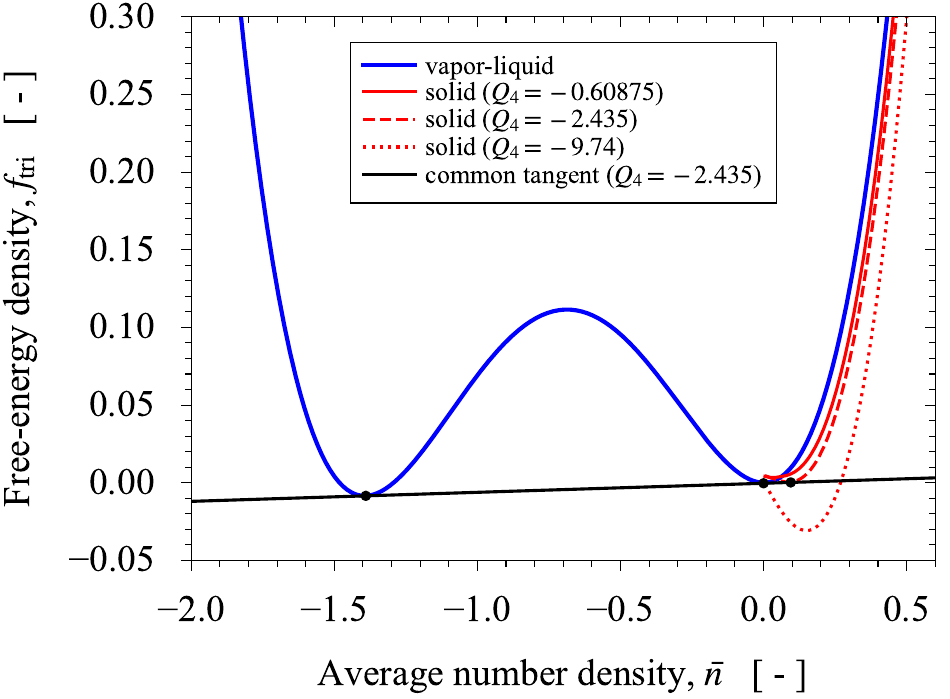}
    \end{tabular}
	\caption{\label{fig:stabilization} Equilibrium free-energy density, $f_\text{tri}(\aven,A)$, in Eq.~\eqref{eq:fetridensity}, considering the amplitude as a function of the average density ($\aven$) and temperature $A=\mathcal{A}[\bar{n},\tau]$, at the triple point temperature $\tau=1.0$ ($T=933$K). The curves correspond to  the uniform (blue) and triangular solid (red) phases as a function of $\aven$, considering the one-mode approximation, for three values of $Q_4$. All other parameters are listed in Table~\ref{tab:modelparams}.}
 \label{vapour_liquid_Hex_free_energy_densities}
\end{figure}

{
We first consider the equilibrium properties of the VXPFC model for the case where $\hat{C}_{2}(\bq)$ given by Eq.~\eqref{basic_XPFC_C2} has a single peak at $|\bq_{10}|$, which will lead to a solid phase with a triangular lattice in two dimensions (2D).} Substituting the three reciprocal lattice vectors $\mathbf{G}=|\bq|(0,1)$ and $|\bq|(\pm\sqrt{3}/2,1/2)$ of a 2D triangular phase into Eq.~\eqref{eq:one-modeapp}, assuming $A_j$ are all equal, yields the following one-mode ansatz for the PFC density field~\cite{Nik_book,wang2020minimal},
\begin{align}
    n(\br_\perp)=
    \aven + A[2\cos(|\bq|y)+4\cos\left(\tfrac{1}{2}|\bq| y\right)\cos\left(\tfrac{\sqrt{3}}{2}|\bq|x\right)] \,,
    \label{eq:triangular}
\end{align}
where $\br_\perp=(x,y)$, $\bar{n}$ is the average density of a phase and $A_j=A$ for all $j$, and $\bq\to\bq_{10}$, in this work. Substituting Eq.~\eqref{eq:triangular} into Eq.~\eqref{eq:vxpfcmodel2} and 
integrating the result over the 2D hexagonal unit cell area yields the equilibrium free-energy density in terms of the variables $\bar{n}$ and $A$, i.e.,
\begin{align}
    f_\text{tri}(\bq,A;\aven,\tau) =& - \tau\big[ B_0 \aven +\tfrac{1}{2}C_0 \aven^2
    +\tfrac{1}{3!}D_0 \aven^3 +\tfrac{1}{4!} E_0\aven^4\big] 
    \nonumber\\&
    -3\tau\big[\big(\hat{\tilde{C}}_{2}(\bq)-1\big)
    + \tfrac{1}{3}\aven\big(3 
    + 2D_1\hat{\tilde{C}}_{3}(\bq) 
    \nonumber\\&
    + D_2\hat{\tilde{C}}_{3}^2(\bq)\big)
    +\tfrac{1}{12}
    \aven^2\big( 3E_1\hat{\tilde{C}}_{4}(\bq)
    \nonumber\\&
    +E_2\hat{\tilde{C}}_{4}^2(\bq)-12
    \big)\big]A^2
    -\tau\big[2\big(
    D_1\hat{\tilde{C}}_{3}(\bq)
    \nonumber\\&
    +D_2\hat{\tilde{C}}_{3}^2(\bq) + 1\big)
    +\tfrac{1}{2}\aven\big(3E_1 \hat{\tilde{C}}_{4}(\bq)
    \nonumber\\&
    +2E_2\hat{\tilde{C}}_{4}^2(\bq)
    +E_3 \hat{\tilde{C}}_{4}^3(\bq)-8\big)\big]A^3
    \nonumber\\&
    -\tfrac{15}{4}\tau\big[E_1 \hat{\tilde{C}}_{4}(\bq)
    +E_2\hat{\tilde{C}}_{4}^2(\bq)
    \nonumber\\&
    +E_3 \hat{\tilde{C}}_{4}^3(\bq)-2\big]A^4\,,
    \label{eq:fetridensity}
\end{align}
where the correlation kernels $\hat{\tilde{C}}_{m}(\bq)$ are kept in a general form, highlighting the connection with other PFC models using higher order correlations designed in real space by expansions in gradient terms (see Appendix~\ref{sec:singlefield}). We note that in the long wavelength limit ($\bq\to0$), coefficients from the excess free energy terms act \textit{only} on the polynomial terms in average density, i.e., not involving the amplitude, $A$. The  coefficients of the correlation kernels $\hat{\tilde{C}}_2$, $\hat{\tilde{C}}_3$ and $\hat{\tilde{C}}_4$ in Eq.~\eqref{eq:fetridensity} are given by Table~\ref{tab:modelparams}. A plot of the free-energy landscape described by Eq.~\eqref{eq:fetridensity} for $\bq=\bq_{10}$, at the triple point ($T=933$K, $\tau=1.0$), is shown in Fig.~\ref{free_energy_landscape_fig}. 
Equation~\eqref{eq:fetridensity} is used to find the free-energy density for the uniform phases (vapor and liquid) by setting $A=0$, which gives
\begin{align}
    f_\text{uniform}(\aven,\tau) = - \tau\big[ B_0 \aven +\tfrac{1}{2}C_0 \aven^2
    +\tfrac{1}{3!}D_0 \aven^3 +\tfrac{1}{4!} E_0\aven^4\big] .
    \label{eq:funiform}
\end{align}
The equilibrium free-energy density of the solid is found by first minimizing Eq.~\eqref{eq:fetridensity} with respect to $A$, i.e. solving $\partial f_{\rm tri}/\partial A=0$, which yields an expression $A=\mathcal{A}[\bar{n},\tau]$. Substituting this back into Eq.~\eqref{eq:fetridensity} yields a function $f_{\rm tri}(\bar{n},\tau)$.
Some typical plots of the equilibrium free-energy densities $f_\text{uniform}$ and $f_{\rm tri}$ are shown in Fig.~\ref{vapour_liquid_Hex_free_energy_densities}, for three values of the $Q_4$ parameter. 

\begin{figure*}[t]
    \begin{tabular}{c}
         \textbf{\includegraphics[width=.975\textwidth]{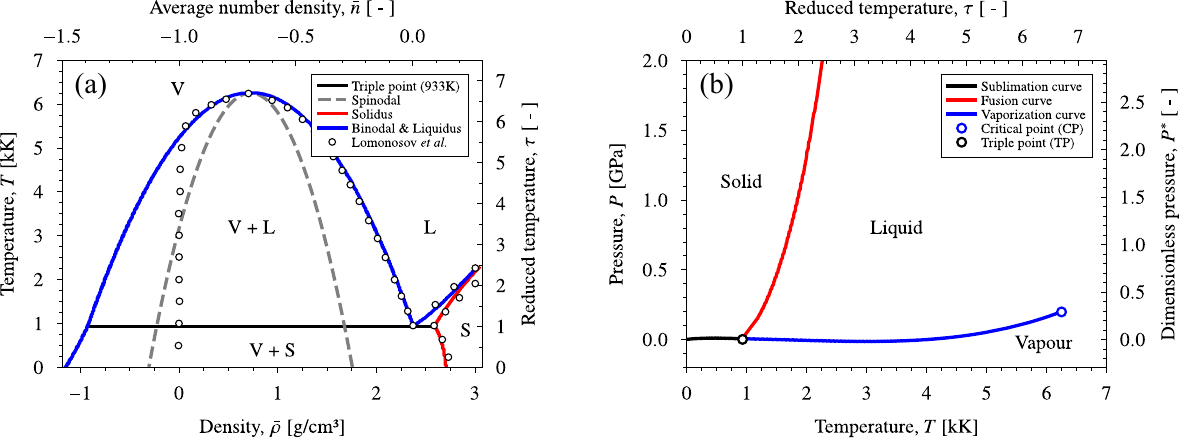}}
    \end{tabular}
	\caption{\label{fig:aluminum} 
    {
    (a) Temperature-density ($T$-$\bar{\rho}$) of the VXPFC model for aluminum (Al), where $\bar{n}=(\bar{\rho}-\rho_0)/\rho_0$. The reference density is taken to be that of  liquid Al at the triple point, $\tilde{\rho}_0=2,368\,\text{kg}\cdot\text{m}^{-3}$. All other parameters are specified in Table~\ref{tab:modelparams}. Solid lines are equilibrium coexistence boundaries (binodal, \textit{liquidus}, and \textit{solidus}). The dashed line is the spinodal curve. The circle dots indicate the temperature-density ($T$-$\bar{\rho}$) phase diagram obtained for aluminum via equation of state (EoS) calculations by Lomonosov \textit{et al.}~\cite{lomonosov2007multi}. The corresponding (b) pressure-temperature ($P$-$T$) phase diagram, where solid lines indicate the sublimation, fusion, and vaporization curves. The critical point and triple point are indicated by blue and black circles, respectively.}}
\end{figure*}

The equilibrium free-energy densities
can be analyzed to find the densities of coexisting phases. These are obtained by applying the  {\it common tangent} rule at each temperature (or {\it Maxwell equal area construction}), which amount to equating the chemical potentials and pressures of coexisting phases ($1$ and $2$) , thus yielding the following equations:
\begin{align}
    \label{eq:commontangent}
    \dfrac{\partial f(\bar{n}_1)}{\bar{n}_1}\bigg\vert_\tau
    = \dfrac{\partial f(\bar{n}_2)}{\bar{n}_2}\bigg\vert_\tau = \mu_\text{eq}
    \,,\nonumber\\
    \mu_\text{eq}=\dfrac{f(\bar{n}_1)-f(\bar{n}_2)}{\bar{n}_1-\bar{n}_2}\,,
\end{align}
where $\mu_{\rm eq}$ is the equilibrium chemical potential, which is equal for both phases. The solution of these three equations gives the coexistence densities, $\aven_1$, and $\aven_2$ for phases 1 and 2, respectively, as well as their common chemical potential $\mu_{\rm eq}$. Applying Eqs.~\eqref{eq:commontangent} to Eq.~\eqref{eq:funiform} yields
\begin{align}
    \aven_{\ell,v}=\bigg(-D_0\pm\sqrt{3D_0^2-6C_0E_0}\bigg)/{E_0} \,,
    \label{eq:eq:commontangentsolution}
\end{align}
where $n_{\ell}$ and $n_{v}$ correspond to the coexistence densities for the liquid and vapor, respectively. The spinodal curve is given by solving $\partial^2 f_\text{uniform}/\partial\aven^2=0$, which gives
\begin{align}
    \aven_\text{spinodal}=\bigg(-D_0\pm\sqrt{D_0^2-2C_0E_0}\bigg)/{E_0}\,.
\end{align}
Applying Eqs.~\ref{eq:commontangent} to $f_{\rm tri}$, in Eq.~\eqref{eq:fetridensity}, similarly gives the solid-liquid phase coexistence densities, which are now solved for numerically. Figure~\ref{fig:aluminum}(a) shows the complete $(\bar{\rho},T)$ phase diagram for the VXPFC model for the case of a triangular (2D) solid phase. 

We also compute the pressure difference of the uniform phases (vapor, liquid, or coexisting vapor and liquid) by considering the case of a fixed total number of particles $N$ and volume $V$ of the system, i.e., $N=\int d\br\rho=\bar{\rho}V=\rho_0(\aven+1)V$, and recalling that $F= \Delta F + f_0\,V$, where, under equilibrium conditions, $\Delta F = (k_BT_0 \rho_0)\,f \, V$ (where $f=f_{\rm uniform}$ with the subscript dropped for ease of notation). This gives,
\begin{align}
    P=-\dfrac{\partial F}{\partial V}\bigg\vert_{T,N}
    =f_0+k_BT_0 \rho_0\left[-f+(\aven+1)\dfrac{\partial f}{\partial\aven}\right] \,,
\end{align}
from which the dimensionless pressure becomes,
\begin{align}
    P^*
    =P_0^*-f+(\aven+1)\dfrac{\partial f}{\partial\aven} \,,
    \label{pstar_uniform}
\end{align}
where $P_0^*={f_0}/{(k_BT_0 \rho_0)}$ is the reference pressure. In this work, it was chosen to be the pressure at the triple point, arbitrarily set to $P_0^\ast= 1.46877\times 10^{-4}$ ($P_0=10$ MPa).  Substituting Eq.~\eqref{eq:funiform} into Eq.~\eqref{pstar_uniform} and evaluating at the density of the liquid corresponding to vapor-liquid coexistence gives the equilibrium pressure difference between  uniform phases,
\begin{align}
    \Delta P_{\text{uniform}}^* =& -\tau B_0-\dfrac{\tau}{2}C_0(\aven^2+2\aven)
    -\dfrac{\tau}{3!}D_0(2\aven^3+3\aven^2)
    \nonumber\\&
    -\dfrac{\tau}{4!}E_0(3\aven^4+4\aven^3) \,.
    \label{eq:pressure}
\end{align}
By proceeding as above, but evaluating Eq.~\eqref{pstar_uniform} at the density of the vapor corresponding to solid-vapor coexistence \textit{and} at the density of the liquid corresponding to solid-liquid coexistence, it yields the respective $P^\ast$ vs. $\tau$ curves for both coexistence lines. Figure~\ref{fig:aluminum}(b) shows the  phase diagram for the solid-liquid-vapor system corresponding to a triangular phase in ($P$, $T$) space, or equivalently, in $(P^*,\tau)$ space.


The above analysis using the one-mode expansion in Eq.~\eqref{eq:one-modeapp} results in a free-energy density for the uniform phases in Eq.~\eqref{eq:funiform} depends on 6 parameters (recalling Eq.~(\ref{eq_B0C0})). The common tangent rule for the liquid-vapor coexistence yields a quadratic function in the parameters ($C_{00}$, $D_0$, and $E_0$). These parameters are obtained by fitting Eq.~\eqref{eq:eq:commontangentsolution} to a parabolic fit of the corresponding part of the real aluminum phase diagram. Further, we use the equilibrium pressure at the critical point ($T_\text{CP}$) to find $B_{00}$, the pressure at the triple point to determine  $B_{01}$, while $C_{01}$ is found by forcing liquid coexistence density along the vapor-liquid coexistence to terminate at the triple point. The above fitting procedure can also be extended to better approximate the temperature-dependent slopes of the phase diagram in $(P, T)$ space using the specific heat and Clausius-Clapeyron relation. Furthermore, these approaches can also be used to calculate the parameters of the solid-liquid and solid-vapor coexistence line in the phase diagram. This will be left for future work. 

We note that the phase diagram for the uniform phases (vapor and liquid) cannot be entirely fit quantitatively due to the limitation of expanding the logarithms in the ideal free-energy term in a Landau-type of polynomial form. This limits the solution for the coexistence lines of the uniform phases to quadratic order in density, which holds quantitatively around the critical point as predicted by mean-field theory, but deviates from the true  \textit{gaseous} curve at lower average densities. This could be circumvented by adding more parameters to the model to better match vapor density below the critical density, although all polynomial expansions would ultimately fail at sufficiently low density. This compromise is done for dynamical stability, and is not expected to impact physical processes such as cavitation, voids, cracks, and surfaces.

\subsection{Solid-liquid-vapor phase coexistence in two dimensions: square phase}

\begin{figure}[h]
	\includegraphics[width=.45\textwidth]{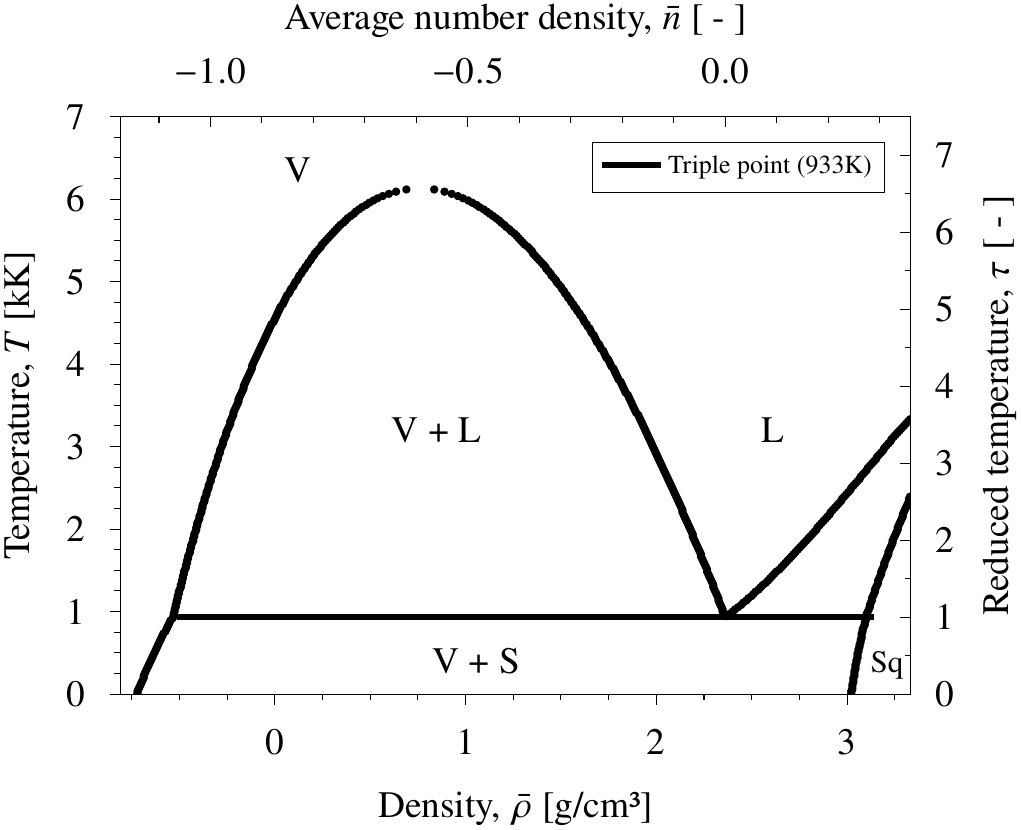}
	\caption{\label{fig:squaresdiag} Temperature-density ($T$-$\bar{\rho}$) phase diagram for the VXPFC model for the case of a solid with a square (2D) lattice.  Model parameters used are listed in Table~\ref{tab:modelparams}.}
\end{figure}

{
This section examines the equilibrium properties of the VXPFC model for the case where $\hat{C}_{2}(\bq)$ defined in Eq.~\eqref{basic_XPFC_C2} contains two peaks at $|\bq_{10}|$ and $|\bq_{11}|$, which will lead to a solid phase with a square lattice in two dimensions (2D).
}To study equilibrium properties of the model for a square solid phase, we expand the density field in Eq.~\eqref{eq:one-modeapp} by substituting the two reciprocal lattice vectors $\mathbf{G}=|\bq|(1,0)$ and $|\bq|(0,1)$ of a 2D square, with different $A_j$. This  yields the following two-mode approximation~\cite{greenwood2011}
\begin{align}
    n(\br_\perp)=&\;
    \aven + A_{10}
    \big[
    (3\sqrt{2}-2)\cos(|\bq|y)-2\cos(|\bq|x)
    \big]
    \nonumber\\&
    -A_{11}\cos(|\bq|x)\cos(|\bq|y) \,,
    \label{eq:squares}
\end{align}
where $\bq\to\bq_{10}$, when generating the phase diagram. Following a similar procedure outlined in the previous subsection for a system with a triangular lattice, the equilibrium free-energy density must now be numerically minimized with respect to two amplitudes $A_{10}$ and $A_{11}$, which then are substituted into the free-energy density of the square lattice. The results in the phase diagram in Fig.~\ref{fig:squaresdiag}, which shows coexistence regions between vapor, liquid, and the square solid phase.

    \section{Dynamical analysis of Model}\label{sec:vxpfcdyn}

    This section explores microstructure dynamics in the VXPFC model. The parameters of the model used in all dynamical simulations are listed in Table~\ref{tab:modelparams}, which were selected to quantitatively match some equilibrium properties of Aluminum, whose phase diagram is shown in Fig.~\ref{fig:aluminum}.  
We first present the time evolution equation governing the density field $n(\br,t)$, which is discretized via the numerical scheme outlined in Appendix~\ref{sec:compmethod}. Secondly, we address the main motivation for developing the VXPFC formalism introduced in Sec.~\ref{sec:model}, namely how interfaces between solid, liquid, and vapor phases are free of dynamical artifacts. Finally, we explore various modalities of microstructure evolution that can emerge during solidification of a pure material driven by the VXPFC model, highlighting that the new model lends itself to the investigation of defects and void formation in rapid solidification processes.

\subsection{Evolution equation of the PFC density field}

As with all PFC models, $n(\mathbf{r},t)$ consists of a reduced number density that must be conserved in a closed system. Its evolution must be governed by the continuity equation, with a flux driven by gradients in the chemical potential, derived from the free-energy functional. This is formally derivable from considerations of the Dynamical Density Functional Theory (DDFT) formulation~\cite{van2009derivation}, and also heuristically derived~\cite{archer2006dynamical}. Following the original works with the PFC framework~\cite{elder2002,Elder2004,elder2007}, the conserved dynamics for the reduced density field, in the limit of constant mobility, becomes
\begin{align}
    \dfrac{\partial n(\mathbf{r},t)}{\partial t}
	=\nabla^2\left(\dfrac{\delta\Delta\mathcal{F}}{\delta n(\mathbf{r})} \right)+ \xi(\mathbf{r},t)\,,
\end{align}
where the noise is a Gaussian stochastic variable satisfying $\langle\xi(\mathbf{r},t)\rangle=0$ and $\langle\xi(\mathbf{r},t)\,\xi(\mathbf{r}^\prime,t^\prime)\rangle = -N_a^2\nabla_c^2\delta(\mathbf{r}-\mathbf{r}^\prime)\delta(t-t^\prime)$, where $N_a^2$ is the amplitude, arbitrarily chosen. We note that $N_a$ typically depends on temperature, although not explicitly defined here. The subscript ``c'' in $\nabla_c^2$ indicates the cutoff around the inter-atomic length scale, necessary for attaining a quantitative match with capillary fluctuation theory~\cite{kocher2015,kocher2016incorporating}, which is important, but not addressed in this work. The computational implementation of the noise (and cutoff) in the simulations is also described in Appendix~\ref{sec:compmethod}. The value of $N_a$ used in all simulations in this section is given in Table~\ref{tab:modelparams}. Taking the first functional derivative of the model in Eq.~\eqref{eq:vxpfcmodel2} yields
\begin{align}
    \label{eq:dynamicalVXPFC2}
    \nonumber
    \dfrac{\partial n}{\partial t}=\;
	\tau\nabla^2\bigg\{ &
	- C_0 n - \tilde{C}_2\ast n
	\\\nonumber&
	- \dfrac{1}{3!}\bigg[
	3D_0n^2 +
    D_1 \bigg(2n\eta_{(3)} + 
    \tilde{C}_3\ast n^2\bigg)
	\bigg]
	\\\nonumber&
    +D_2
	\bigg(\eta_{(3)}^2 + 2\tilde{C}_3\ast(n\eta_{(3)})\bigg)
	\\\nonumber&
	- \dfrac{1}{4!}\bigg[
	4 E_0 n^3
    +E_1\bigg(3n^2\eta_{(4)} + \tilde{C}_4\ast n^3\bigg)
	\\\nonumber&
    +E_2\bigg(2n\eta_{(4)}^2 + 2\tilde{C}_4\ast(n^2\eta_{(4)})\bigg)
    \\&
    +E_3\bigg(\eta_{(4)}^3 + 3\tilde{C}_4\ast(n\eta_{(4)}^2)\bigg)
	\bigg]
	\bigg\} + \xi(\mathbf{r},t) \,,
\end{align}
where $\eta_{(m)}=\{C_m\ast n\}(\br)$ to be computed as the inverse Fourier transform $[\hat{C}_m(\bq)\hat{n}_\bq]_\br$, and it is understood that in dynamics $n\equiv n(\br,t)$.

{
The following subsections examine different applications of Eq.~\eqref{eq:dynamicalVXPFC2} to interfaces and microstructure evolution. 
It is noted that Eq.~\eqref{eq:dynamicalVXPFC2} can also be utilized in tandem with recent efficient numerical integration schemes for PFC modeling of dislocation dynamics, ballistics, and thermal response in complex material structures \cite{Burns2023, Burns2023_2}.}

\subsection{Interfacial energy properties of model}

We additionally conducted dynamical simulations to study the interfacial energy of solid-vapor, and liquid-vapor interfaces using the VXPFC model. In particular, we examined the robustness of our model to independently control interface energy between different phases. Solid-liquid interfaces behave similarly in this model as in other PFC models that have analyzed such interfaces, and thus will not be discussed here. Each simulation is initially set up analogously to Fig.~(\ref{fig_mean_field}), with arbitrary fractions of solid, liquid, and vapor at their coexistence average densities ($\aven$), and dynamically evolved based on Eq.~\eqref{eq:dynamicalVXPFC2}. The initially sharp interface profiles seeded between phases evolve into smooth profiles by following the dynamical relaxation of the density field, yielding a particular interface width for late times. 

The parameters $\alpha$ and $\beta$ in the correlation functions in Eq.~\eqref{basic_XPFC_C2} are expected to control the interface energy of uniform-solid phases and liquid-vapor phases, respectively. Straightforwardly, one can analytically derive a theoretical prediction to the lowest order in $\bq$ from the model's free energy in Eq.~\eqref{eq:vxpfcmodel2} by expanding separately around $(|\bq|_{i}^2 - |\bq|^2) \rightarrow 0$ and $|\bq| \rightarrow 0$ and a subsequent integration by parts. This is expected to contain up to two contributions to the interface energy, one of the form $\int d\br\,\beta^{-1}|\nabla\aven|^2$, arising from variations of the average density. The other, of the form $\int d\br\,\alpha^{-1}|\nabla A|^2$, where $A$ is the amplitude of the solid phase in the lowest order mode expansion. 

\begin{figure}[t]
	\includegraphics[width=.475\textwidth]{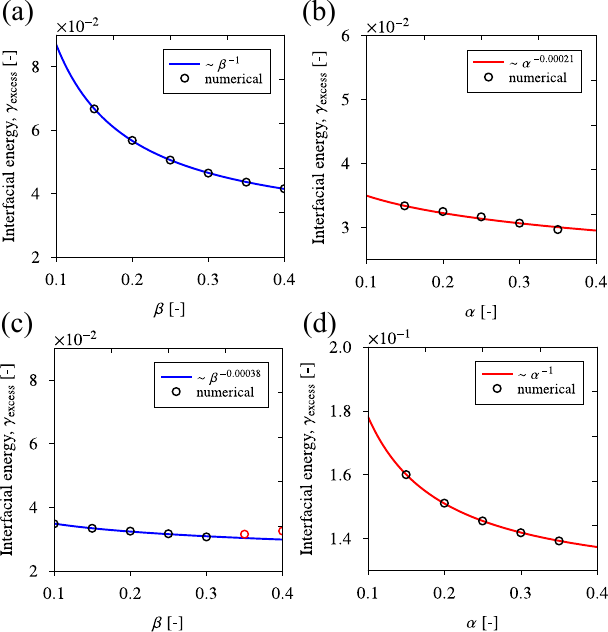}
	\caption{\label{fig_vapour_liquid_int}
    Interface energy as a function of $\beta\in[0.1,0.4]$, and $\alpha\in[0.1,0.4]$ for each of the following cases: (a-b) liquid-vapor, and (c-d) solid-vapor interfaces. An inverse power law is shown for comparison in each case: (a) $0.00606\beta^{-1} + 0.02630$, (b) $18.41639\alpha^{-0.00021} -18.39050$, (c) $9.49633\beta^{-0.00038} -9.46985$, and (d) $0.00545\alpha^{-1} + 0.12367$. In (c), the range $0.35 <\beta <0.4$ corresponds to a pure material whose average density is shifted from values from the phase diagram in Fig.(\ref{fig:aluminum}). We observe that the excess energy associated with vapor-liquid interfaces is inversely proportional to $\beta$, such that $\text{interfacial energy}\sim \int d\br\beta^{-1}|\nabla\aven|^2$, while  $\sim \int d\br\alpha^{-1}|\nabla A|^2$ for interfaces between solid and a uniform phase (liquid or vapor).}
\end{figure}

The numerically calculated interfacial energies for liquid-vapor and solid-vapor interfaces are shown in Fig.~\ref{fig_vapour_liquid_int}. Interface energy was calculated via the dimensionless excess free energy ($\gamma_\text{excess}$), following the procedure in Appendix~\ref{sec:compmethod}. The behavior of the interface energy is examined for several values of $\beta$ in the left-hand columns of the figure. Meanwhile, the behavior of the interface energy is examined for several values of $\alpha$ in the right-hand columns of the figure.

In Fig.~\ref{fig_vapour_liquid_int}(a), we observe that the variation of the liquid-vapor excess energy  with $\beta$ fits well to a power-law form $\gamma_{\rm excess}=0.00606\beta^{-1} + 0.02630$ with a $R^2$ value of $0.999$, which is as expected. Conversely, Fig.~\ref{fig_vapour_liquid_int}(b) shows that the variation of the excess energy of the liquid-vapor interface with $\alpha$ is negligible, which is also as anticipated since the liquid-vapor interface energy is expected to be dominated by long-wavelength variations, controlled by $\beta$. In Fig.~\ref{fig_vapour_liquid_int}(c), the variation of the excess energy of the solid-vapor interface with $\beta$ is very minor, which is also anticipated, because the solid-vapor interface energy is expected to be dominated by amplitude variations, controllable via $\alpha$, separately shown in Fig.~\ref{fig_vapour_liquid_int}(d), which fits well to a power-law form $\gamma_{\rm excess}=0.00545\alpha^{-1} + 0.12367$ with a $R^2$ value of $0.999$. 


The parameter $\beta$ can also be used to explore other types of phase diagrams where the solid phase can be shifted in average density relative to the one shown in Fig.~\ref{fig:aluminum}. For example, in the range $0.3 < \beta < 0.4$ (red circles in Fig.~\ref{fig_vapour_liquid_int}(c)), one produces a shift in the solid region of the phase diagram of a pure material. For $\beta > 0.4$, one obtains a phase diagram corresponding to a material with anomalous density changes, where the equilibrium density of the solid phase lies between that of the liquid and vapor phases~\cite{jreidini2021}. It also noted that the parameter $\beta$ in the VXPFC model plays an analogous role as the ``smoothing'' parameter $\lambda$ in previous models that couple the density to its mean-field ($n_{mf}$). 
As a result, for low values of $\beta$ ($\beta < 0.1$), the correlation function starts to over-smooth the density field, which may lead to the type of artifacts discussed above. However, we expect that the range of $\beta$ shown is adequate to cover a wide range of the phase space of a pure material.

\subsection{Comparison of solid-vapor interfaces to previous models}

In this section we examine the dynamics of solid-vapor interfaces, specifically addressing an artifact reported to dynamically occur at such interfaces simulated by the model of Kocher \textit{et al.}~\cite{kocher2015}, for specific density and temperature ranges. Kocher \textit{et al.}~\cite{kocher2015} introduced couplings of the PFC density field with powers of its mean field in the free energy of the model (as in Eq.~\eqref{eq:kochermodel}), an approach similar to that examined in Refs.~\cite{Emilythesis,jreidini2021,jreidini2022}. In these approaches, these couplings (effectively excess terms) multiplied $n(\br)$ in the free energy by powers of the expression $n_{mf}(\mathbf{r})=\int d\br' \chi(\br-\br')n(\br')$, where the correlation kernel $\chi(\br-\br')$ is designed to operate only on long wavelengths. The expression $n_{mf}(\mathbf{r})$ can also be seen to be a smoothing of the microscopic density $n(\br)$.

\begin{figure}[t]
	\includegraphics[width=.475\textwidth]{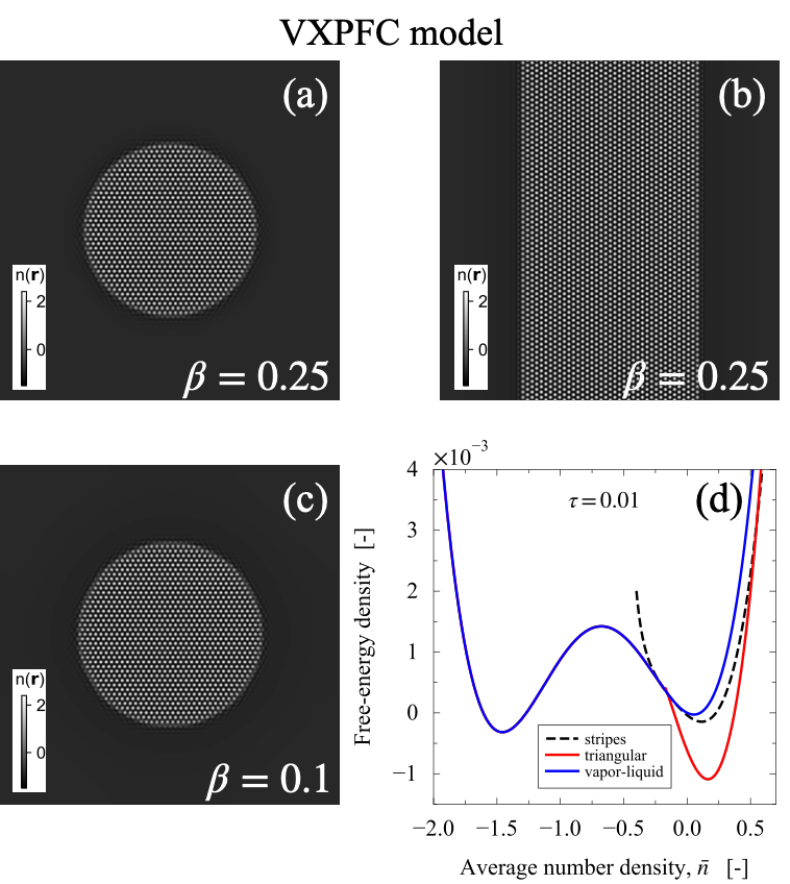}
	\caption{\label{fig_mean_field}
    Nonequilibrium coexistence of solid and vapor phases in two-dimensional (2D) systems simulated using the VXPFC model in Eq.~\eqref{eq:vxpfcmodel2} with the dynamics in Eq.~\eqref{eq:dynamicalVXPFC2}. All simulations were initialized with a system containing a crystal seed and average density set to $\aven=0.05$, and quenched below the triple point at the scaled temperature $\tau =0.01$ following the corresponding phase diagram in Fig.~\ref{fig:aluminum} and parameters in Tab.~\ref{tab:modelparams}. The smoothing parameter is set to $\beta=0.25$ in the (a) and (b) cases, and $\beta=0.1$ in (c). We have used a time step $\Delta t=0.01$, and $N=512^2$ grid points. All cases (a), (b), and (c) do not present any artifacts at the solid-vapor interface. The free energy density curves for triangular, stripes, and liquid phases are shown in (d) for reference.
    }
\end{figure}

{
In the above models, certain artifacts around solid-vapor interfaces that developed during dynamical simulations of sublimation or vaporization of solids was observed.  Fig.~\ref{wilson_vapour_1} shows three examples of this dynamical artifact, which we simulated here using the model and dynamics used by Kocher \textit{et al.} in Eq.~\eqref{eq:kochermodel}.  In particular, Fig.~\ref{wilson_vapour_1}(a) shows a snapshot in the time evolution of an initially circular solid crystal growing into a vapour, while Fig.~\ref{wilson_vapour_1}(b) shows an essentially stabilized slab of solid coexisting with its vapour. In both cases, the quenches are at a model temperature ($\Delta B$) below the triple point, with the other parameters indicated in the figure caption. These images show that there is a sharp beading in the crystal structure at the solid-vapor interface. It was hypothesized in Ref.\cite{frick2023consistent} to be caused by the coupling of the microscopic density field to its mean field, which filters out wavelengths shorter than some cutoff, thus limiting dynamical access to a wide range of wavelengths. We further examined this here by examining the free energy landscape of this model at different model temperatures. We found that this artifact is, in fact, due to the presence of a stripe phase of lower energy which sits very close to the solid density. An example of this is shown in Fig.~\ref{wilson_vapour_1}(d). The proximity in density of the stripe and solid phase makes it easy to spawn a stripe phase in the higher-energy region of the vapour-solid interface,  activated by inevitable numerical fluctuations. To further support this hypothesis, we simulated the growth of a solid crystal growing in its vapour, shown in Fig.~\ref{wilson_vapour_1}(c), using a smaller value of the smoothing parameter $\lambda$ (discussed in the Appendix~\ref{sec:singlefield}). This value of $\lambda$ comes closer to the limit of validity of the simplified dynamics employed by Kocher \textit{et al.} in Ref.~\cite{kocher2015}. Here we see that a more complete looking stripe phase grows into the hexagonal crystal. We observed that the width of this striped regions becomes larger as the smoothing parameter $\lambda$ is further decreased. 
\begin{figure}[t]
	\includegraphics[width=.475\textwidth]{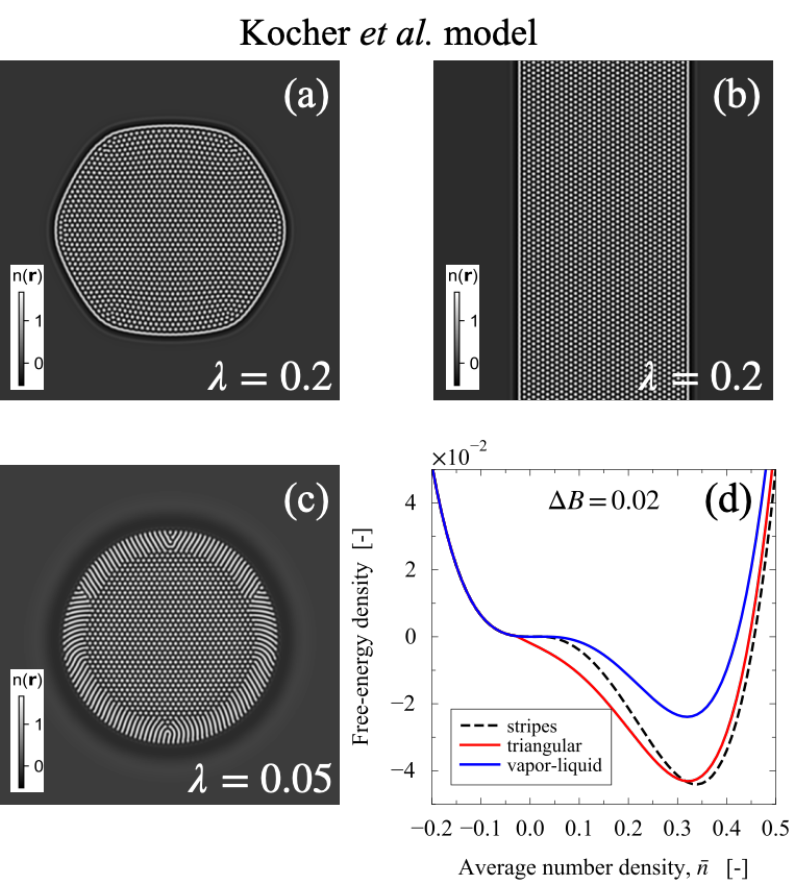}
	\caption{\label{wilson_vapour_1}
    Nonequilibrium coexistence of solid and vapor phases in two-dimensional (2D) systems simulated using the previous vapor-forming PFC model in Eq.~\eqref{eq:kochermodel} following the conserved dynamics described in the Supplemental Material~\cite{SuppZF2018} of Ref.~\cite{kocher2015}. All simulations were initialized with a system containing a crystal seed and average density set to $\aven=0.2$, and quenched below the triple point at the effective temperature $\Delta B=0.02$ following the corresponding phase diagram in Ref.~\cite{kocher2015,frick2023consistent} with this choice of parameters: $B_x=0.3$, $a=33.5$, $b=-12.01$, and $c=35$. The smoothing parameter is set to $\lambda=0.2$ in (a) and (b) cases, and $\lambda=0.05$ in (c). We have used a time step $\Delta t=0.1$, and $N=512^2$ grid points. The solid-vapor interface in cases (a) and (b) features a ``single stripe'' region, while in (b) a larger stripe phase bulk forms.
    The free energy density curves for triangular, stripes, and liquid phases are shown in (d) for reference.
    }
\end{figure}
As a comparison, we used the VXPFC model introduced in this work to simulate the similar vapor-solid structures as shown in Fig.~\ref{wilson_vapour_1}. These results are shown in Fig.~\ref{fig_mean_field} for a quench similarly below the triple point of the phase diagram in Fig.~\ref{fig:aluminum} as it was for the phase diagram of Ref.~\cite{kocher2015}. The VXPFC model parameters used are shown in the  caption of Fig.~\ref{fig_mean_field}. The VXPFC density field was relaxed in all cases following the dynamics in Eq.~\eqref{eq:dynamicalVXPFC2}. As shown in all cases of Figure~\ref{fig_mean_field}. The VXPFC simulations do not produce any beading or striping artifacts. An examination of the free energy curves of the triangular and stripe phases in  Fig.~\ref{fig_mean_field}(d) further reveals that
the stripe phase is well above the energy of the triangular phase, thus making it difficult, if not impossible, to trigger a stripe phase, or beading --a manifestation of the stripe phase- anywhere in our solid crystal. This was also the case for other densities and quench temperature in the regions of the phase diagram of regions of the VXPFC phase diagram in Fig.~\ref{fig:aluminum}.

The above discussion reveals that the ``artifacts'' in the model of Kocher \textit{et al.} \cite{kocher2015}, and its derivatives~\cite{Emilythesis,jreidini2021,jreidini2022} are fundamentally caused by the model having a stripe phase --as do most PFC model-- that cannot be conveniently moved far enough away in density space from the triangular phase of interest. Analogously to the work of Wang \textit{et al.}~\cite{wang2020minimal}, one of the innovations of our model, is it use of higher order correlations, which allow us to more easily manipulate the depth of the solid phase energy, which effectively allows us to place the stripe phase conveniently far from solid phases of interest.}

\subsection{Crystallization from phase separating fluid}

\begin{figure}[t]
	\includegraphics[width=.475\textwidth]{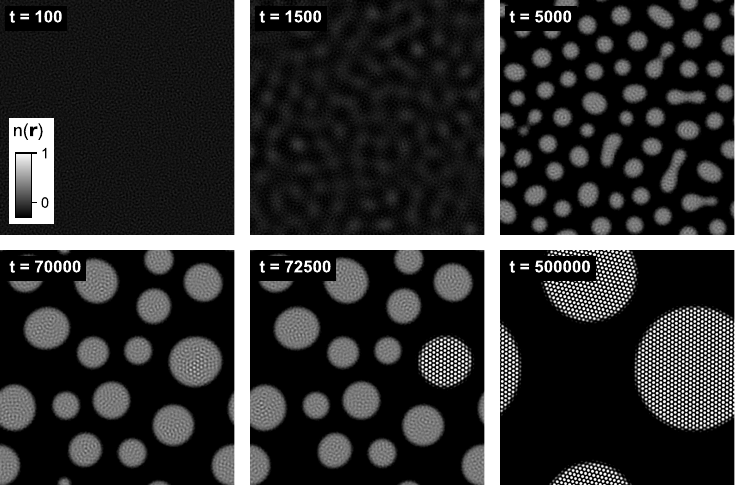}
	\caption{\label{fig:dynamics00}
	The system initially undergoes phase separation into vapor (black) and metastable liquid (gray) phases emergent from the initial supercritical fluid $\aven=-0.85$ ($\bar{\rho}=0.3552\,\text{g}/\text{cm}^{3}$), quenched to $\tau=0.32154$ ($T=300$K). We use $N=1024^2$ and system parameters as in Table~\ref{tab:modelparams} phase diagram in Fig.~\ref{fig:aluminum}. After $t/\Delta t=72000$ time steps, a crystal  nucleate from the metastable liquid drops, with the formation of the first full solid crystal at $t/\Delta t=72500$. Coarsening dynamics follows into late times until two larger circular-shaped (due to surface energy minimization) crystal grains remain by $t/\Delta t=500000$. Further coarsening would result in a single solid mass remaining.}
\end{figure}

The VXPFC model lends itself to study phase separation between liquid and vapor (uniform phases). In this limit, one can regard the model as a classic ``phase field" model whose order parameter is  density difference. 
However, because the order parameter is naturally capable of ordering, it is also possible to follow phase transitions wherein the vapor or liquid can nucleate crystalline phases, the latter of which can exhibit elasto-plasticity. As a demonstration of this, we studied a system that was initialized with a supercritical fluid with an average density of $\aven=-0.85$ ($\bar{\rho}=0.3552\,\text{g}/\text{cm}^{3}$) cooled to $\tau=0.32154$ ($T=300$K), below the triple point. Equation~(\ref{eq:dynamicalVXPFC2}) was then used to follow the subsequent microstructure evolution, based on the parameters in Table~\ref{tab:modelparams} { and the correlation function used to make the triangular phase in Fig.~\ref{fig:aluminum}}. Snapshots of the system evolution are illustrated in Fig.~\ref{fig:dynamics00}. At early times, the metastable fluid nucleates metastable liquid (gray regions) drops that grow into and deplete the surrounding matrix of vapor (black regions).  After $t/\Delta t=72000$ time steps, crystal nucleation and growth is observed within the metastable liquid drops. At later times ($\sim t/\Delta t=500000$), the system is essentially full of coarsening crystal grains in a vapor matrix, which continue to grow more slowly based on surface energy minimization. Further coarsening would result in a single solid mass, with the amount of vapor and solid phases expected to follow the lever rule on the phase diagram, i.e., larger volume fraction of vapor than liquid.
The results of Fig.~(\ref{fig:dynamics00}) are the pure-material analogue of those obtained in~\cite{Smith2017}, which examines crystallization from a phase separating binary alloy in order to explain neutron scattering experiments of this phenomenon \cite{Loh2017}, which we also expect to occur in pure materials. We also note that quenching into liquid-vapor coexistence region close to the critical density in Fig.~\ref{fig:aluminum} would lead to spinodal decomposition into liquid-vapor domains, as described by, e.g., the Cahn-Hilliard equation. This case was studied by Frick \textit{et al.}~\cite{frick2023consistent} using a recent {\it two-field} solid-liquid-vapor PFC model. This is in contrast to the example in Fig.~\ref{fig:dynamics00} examined herein, where the system rapidly evolves to a configuration of many small droplets via nucleation, which subsequently also undergo Ostwald ripening at late times. 
\begin{figure}[tbh]
	\includegraphics[width=0.475\textwidth]{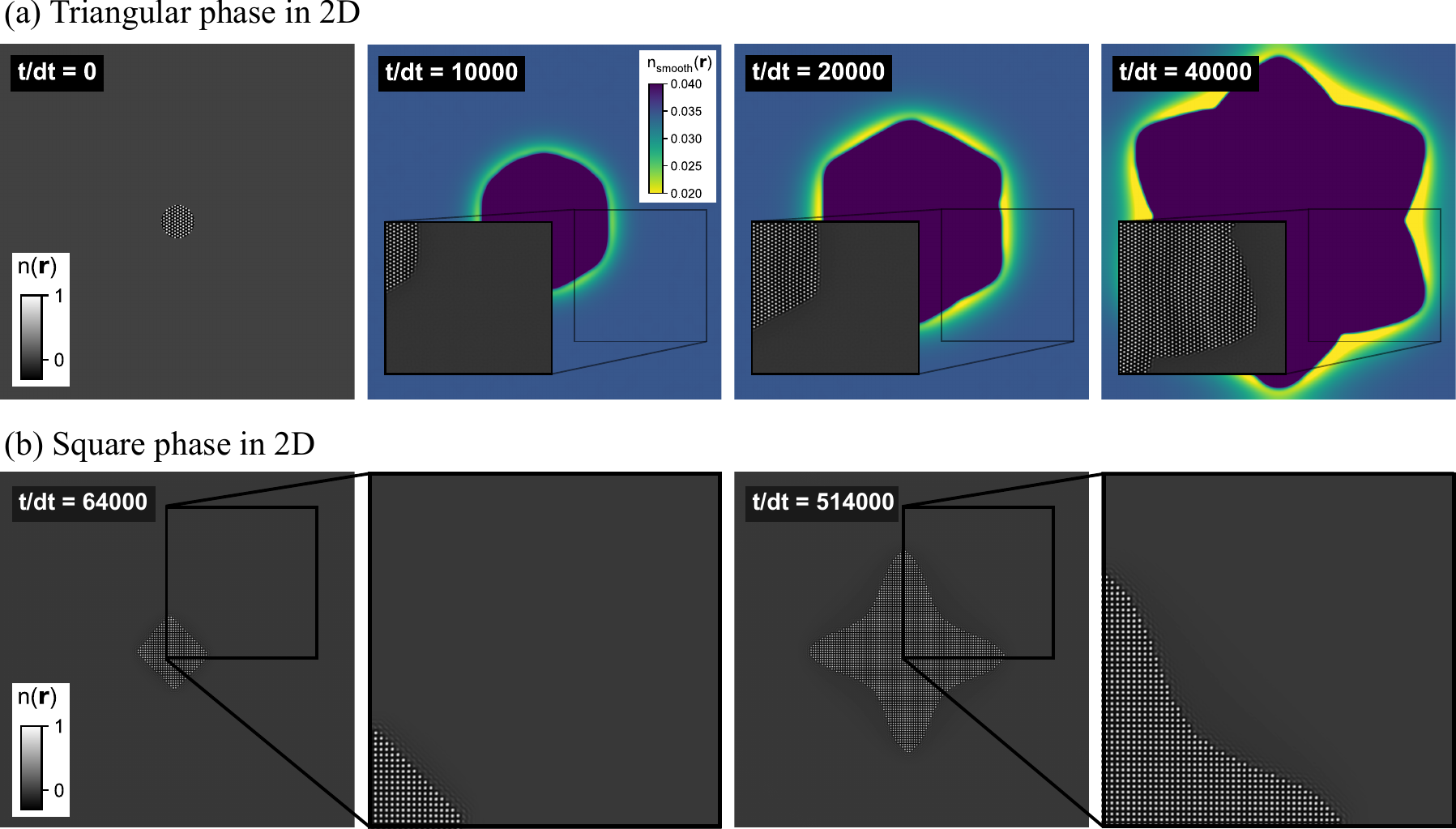}
	\caption{\label{fig:dynamics04323} 
    Two-dimensional (2D) anisotropic crystal growth of dendritic crystals into an undercooled melt. The first row labelled by (a) corresponds to the triangular solid phase growth. The system is initialized at $t/\Delta t=0$ (first frame) with a crystal seed  surrounded by undercooled liquid (grey) at $\aven=0.1$ ($\bar{\rho}=2.6048\,\text{g}/\text{cm}^{3}$), and quenched to $\tau=1.286$ ($T=1200$K) in the phase diagram in Fig.~\ref{fig:aluminum}. The atomic density field $n(\br)$ is shown in gray scale. The subsequent time steps show the averaged density field, smoothed over a unit cell. A section of the atomic density field around a dendritic arm is enlarged in the inset of each frame. The second row labelled by (b) corresponds to the growth of a solid with a square lattice. Growth occurs from an undercooled liquid (grey) at $\aven=0.2$ ($\bar{\rho}=2.8415\,\text{g}/\text{cm}^{3}$) and quenched to $\tau=1.608$ ($T=1600$K) in the phase diagram in Fig.~\ref{fig:squaresdiag}. For both simulations, we used $N=1024^2$.
       }
\end{figure}

\subsection{Dendritic solidification from an undercooled melt}

In order to probe the solidification kinetics at fixed volume with the VXPFC model, we perform two simulations of anisotropic crystal growth of dendritic crystals into an undercooled melt in 2D.  This is
demonstrated in Fig.~\ref{fig:dynamics04323}(a), corresponding to growth of a solid with a { triangular lattice} produced by the single dominant high-$q$ peak in $\hat{C}_2(\bq)$ as in Fig.~\ref{fig:2PCF}(a). The first frame shows the system initialized with a crystal seed surrounded by undercooled liquid (gray) at $\aven=0.1$ ($\bar{\rho}=2.6048\,\text{g}/\text{cm}^{3}$), and quenched to $\tau=1.286$ ($T=1200$K), according to the phase diagram indicated in Fig.~\ref{fig:aluminum}. The subsequent frames show the averaged atomic density, smoothed over a unit cell according to $n_{\text{smooth}}(\br) = \big[\exp(-q^2 / 0.1^2)\hat{n}_\bq\big]_\br$, where $[\cdot]_\br$ represents the inverse Fourier transform, and $\hat{n}_\bq$ is the Fourier transform of $n(\br)$. 
A section of the atomic density field around the dendritic arm is also shown in the insets. After an initial transient time, the spherical morphology becomes unstable, and the solid shape begins to express the preferred growth directions of the underlying crystal due to the surface energy anisotropy of the six-fold triangular lattice.

{Figure~\ref{fig:dynamics04323}(b) shows a simulation of the dendritic growth of a solid phase with a square lattice symmetry (two dominant high-$q$ peaks in $\hat{C}(\bq)$) as in Fig.~\ref{fig:2PCF}(b)}. An undercooled liquid (gray) at $\aven=0.2$ ($\bar{\rho}=2.8415\,\text{g}/\text{cm}^{3}$) is quenched to $\tau=1.608$ ($T=1600$K), in the phase diagram in Fig.~\ref{fig:squaresdiag}.  After an initial transient, the spherical morphology becomes unstable, and the solid shape begins to express the preferred growth directions of the underlying crystal due to the surface energy anisotropy of the four-fold symmetry of the square lattice.

\begin{figure}[t]
    \begin{tabular}{c}      \includegraphics[width=.475\textwidth]{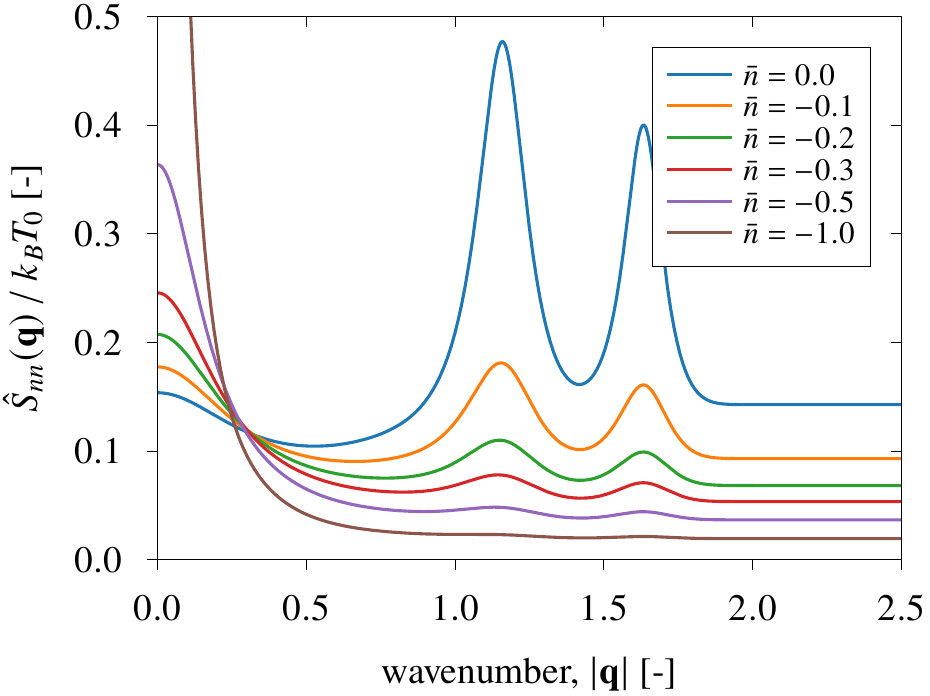}
    \end{tabular}
	\caption{\label{fig:ursell}
    The Ursell function, Eq.~\eqref{eq:ursell}, derived in Appendix~\ref{sec:strucfactor} for different values of the average density $\bar{n}$, using a two-point direct correlation $\hat{C}_2(\bq)$ with two high-$q$ peaks, which support the emergence of both triangular and square phases in two dimensions (2D). The parameters used are listed in Table~\ref{tab:modelparams}.}
\end{figure}

\begin{figure}[tbh]
	\includegraphics[width=\columnwidth]{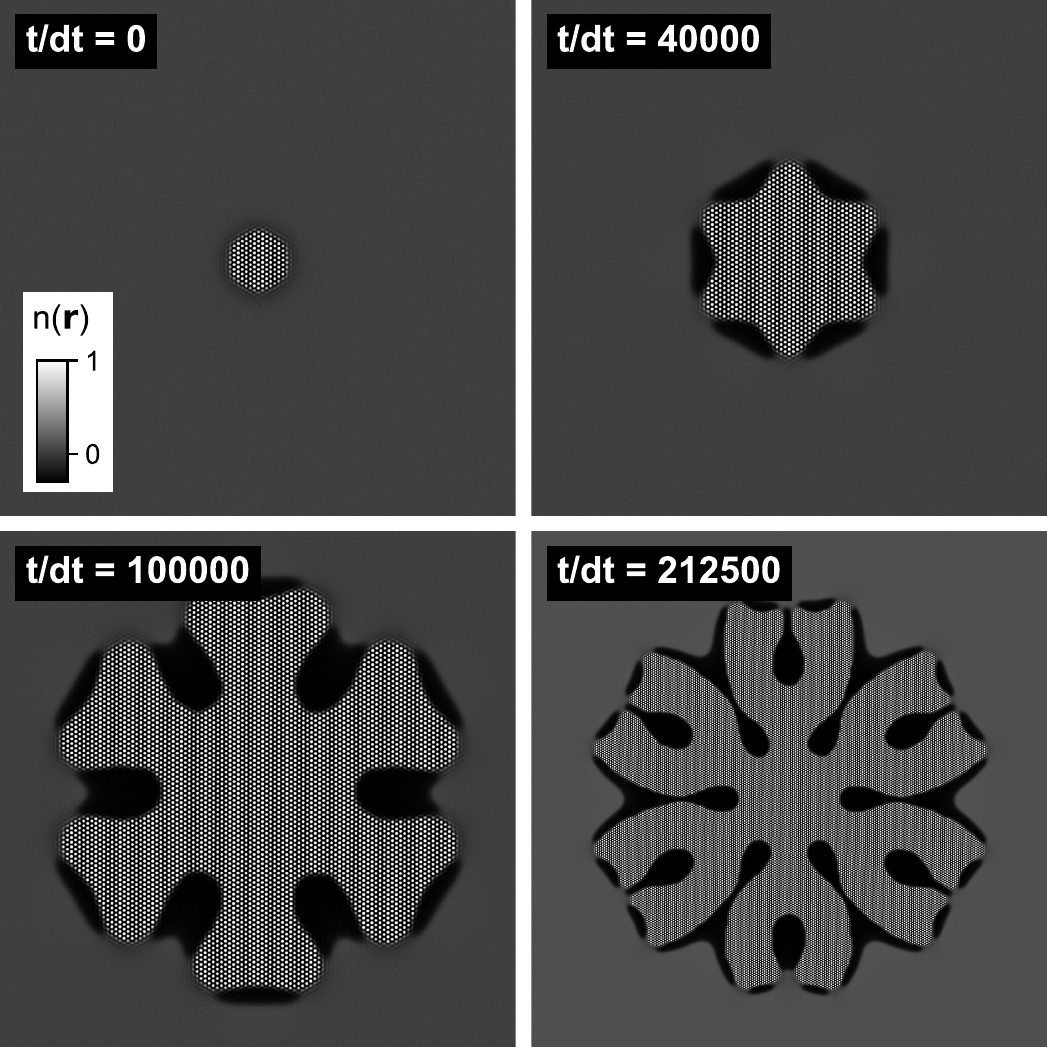}
	\caption{\label{fig:dynamics04} 
    Two-dimensional (2D) seaweed growth into an undercooled melt through vapor (black) deposition from depletion zones formed in a meta-stable liquid (gray). A uniform liquid at $\aven=0.02$ ($\bar{\rho}=2.4154\,\text{g}/\text{cm}^{3}$) was quenched to $\tau=0.6431$ ($T=600$K), i.e., into solid-vapor coexistence in the phase diagram in Fig.~\ref{fig:aluminum}. The metastable liquid is seeded with a solid crystal that grows with a six-fold symmetry of a triangular lattice. As the solid depletes the surrounding liquid density, vapor pockets nucleate in the high depletion areas, driving further dendritic growth, which forms a seaweed-like structure at late times. We use $N=1024^2$.
}
\end{figure}

We also examined dendritic growth, involving kinetics between all three phases. This is shown in Fig.~\ref{fig:dynamics04}. In this case, a uniform liquid at $\aven=0.02$ ($\bar{\rho}=2.4154\,\text{g}/\text{cm}^{3}$) was quenched to $\tau=0.6431$ ($T=600$K), i.e., into the solid-vapor coexistence of the phase diagram indicated in Fig.~\ref{fig:aluminum}, and a solid crystal (triangular phase) is seeded in the metastable liquid. As the surrounding liquid density is depleted by the growing solid, vapor pockets nucleate in high depletion areas. This then further drives the formation of the crystal phase. Rapid solidification kinetics leads to characteristic tip-splitting of the initial six-fold crystal, that lead to seaweed-like dendrite, analogously to what is seen in the stability of doublon formation \cite{Doublon_2004}. This type of dendritic growth via vapor deposition has been also examined by~\cite{kocher2015,frick2023consistent} using a previous generation of vapor-forming PFC model described in the introduction, and shown in Appendix~(\ref{sec:singlefield}) to be a special case of the VXPFC model proposed in this work.

{
We close this subsection by illustrating the transition of the VXPFC model from the long wavelength ``phase field'' limit at low densities, where liquid and vapor phases are described by a smooth order parameter, to the PFC limit where the order parameter becomes periodic to support crystalline phases. We perform a linear stability analysis of the model around a uniform fluid state of average density $\aven$, and the two peak correlation kernels to characterize the solid phase, from which one can generate a triangular or square phase. The details of the calculation are outlined in Appendix~\ref{sec:strucfactor}. We derive the corresponding Ursell function, $\hat{S}_{nn}(\bq)$, which is plotted in Fig.~\ref{fig:ursell} for various values of $\bar{n}$. As the average density of the uniform fluid state decreases, i.e., approaches values closer to the vapor densities, the Ursell function behaves, to the lowest order, as $\hat{S}_{nn}(\bq)\sim \hat{S}_{nn}(0) -|\bq|^2\,$. This indicates that the total direct correlation function of the model, $\hat{C}_{nn}(\bq)=[1-\hat{S}_{nn}(\bq)]/\hat{S}_{nn}(\bq)$ does not feature prominent contributions from high-$q$ peaks. Conversely, as the average density of a fluid increases, the rise of high-$q$ peaks signals the emergence of an atomically ordered crystal phase.}

\subsection{Void formation at grain boundaries}

One of the main motivations and interest in the development of the VXPFC model is the ability to simulate cavitation and void formation in metals and their alloys, which are key defect mechanisms in processes ranging from additive manufacturing to Hydrogen Embrittlement (HE). It is typically during manufacturing that void nucleation, growth, and coalescence take place. Voids are formed by the accumulation of vacancies through bulk diffusion, which may be driven by residual stress gradients or simply to reduce the surface free energy. Voids have been shown to initiate crack at grain boundaries (GBs) that drive intergranular fracture in polycrystalline materials~\cite{moser2021electropolishing,konishi2002effect}. This failure mechanism can be enhanced by the presence of non-metallic impurities, such as hydrogen (H), which tends to segregate at grain boundaries and form voids, thus negatively affecting mechanical performance of metals~\cite{chene2004role}. Recent work by  Fotopoulos \textit{et al.} has also examined this reduction in the ductility (yield stress) of a metal due to absorbed H in bicrystalline Cu films using theoretical calculations (DFT) and bond-order potential Molecular Dynamics (MD) simulations~\cite{fotopoulos2023simulation,fotopoulos2023structure}. We expect that future VXPFC modeling can compliment such studies to elucidate this phenomenon over much longer time scales and on larger system domains.

\begin{figure*}
	\includegraphics[width=\textwidth]{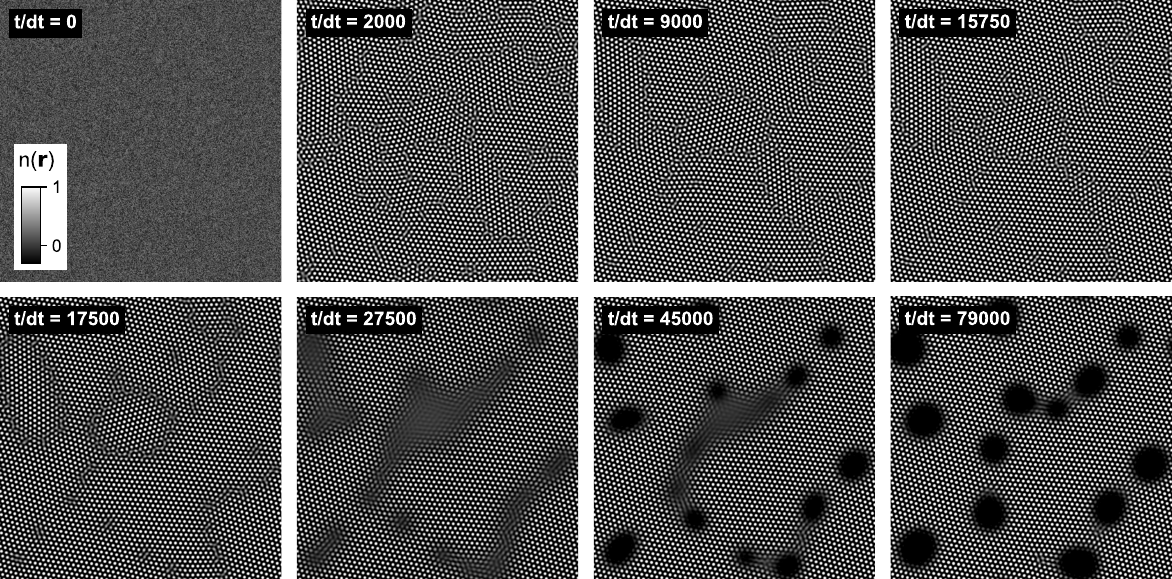}
	\caption{\label{fig:voidnucleation} 
    Nucleation and growth of voids at the grain boundaries of a polycrystalline material simulated by the VXPFC model in two dimensions. The first row (a) displays the evolution of a polycrystal (triangular symmetry) from a noise-perturbed liquid state (gray) at $\aven=0.2$ ($\bar{\rho}=2.8416\,\text{g}/\text{cm}^{3}$) and quenched to $\tau=0.322$ ($T=300$K) in the phase diagram in Fig.~\ref{fig:aluminum}. The atomic density field $n(\br)$ is represented in gray scale as shown in the $t/\Delta t=0$ frame. The second row (b) shows nucleation and growth of vapor pockets at the grain boundaries of the  polycrystalline sample in (a). The system in (b) is initialized from the microstructure at the time step $t/\Delta t=15750$ in row (a), but had its average density shifted to $\aven=-0.2$ ($\bar{\rho}=1.8944\,\text{g}/\text{cm}^{3}$), at the same temperature, thus placing it into the solid-vapor coexistence region of the phase diagram. This was done as a simple way to demonstrate the placement of a polycrystalline sample in a vapor-rich environment. From time step $t/\Delta t=17500$ to $t/\Delta t=27500$, grain boundary (GB) premelting starts to occur, since the free energy of solid–liquid interfaces is lower than that of a dry GB. The formation of metastable liquid pools comprises an intermediate step for the subsequent nucleation of vapor pockets (black) at locations near high energy crystalline defects. We use $N=512^2$ and parameters as in Table~\ref{tab:modelparams}.}
\end{figure*}

 In PFC modeling, it was recently shown that vacancies can be represented by changes in the local density amplitude \cite{Wang2016,Burns2022}. We thus expect that the VXPFC type of model can fully capture the interplay of vacancy and void formation.  We examined the phenomenon of void nucleation at grains using the VXFPC model by first quenching { a liquid into the triangular solid region on the phase diagram in Fig.~\ref{fig:voidnucleation}(a)}, where $\aven=0.2$ ($\bar{\rho}=2.8416\,\text{g}/\text{cm}^{3}$) and $\tau=0.322$ ($T=300$K). This leads to a  polycrystalline structure (triangular symmetry) containing dry grain boundaries by time  $t/\Delta t=15750$.  In order emulate the placement of a polycrystalline sample in a vapor-rich environment (as simple demonstration, only, in this work), we then uniformly subtract average density directly from the density field $n(\br)$, which corresponds to shifting the average density of the sample to $\aven=-0.2$ ($\bar{\rho}=1.8944\,\text{g}/\text{cm}^{3}$), while maintaining the same temperature, i.e., the system is placed into solid-vapor coexistence. A metastable intermediary liquid phase forms preferentially at sites of high energy (e.g., GBs) seen between $t/\Delta t=17500$ to $t/\Delta t=27500$, and then phase transforms into the stable vapor phase.

    \section{Conclusion}\label{sec:conclusion}

    We have extended the XPFC modeling formalism to include high-order (three- and four-point) direct correlation functions to study transitions and coexistence among vapor, liquid, and crystalline solid phases within a framework of a single continuum density-field description. The new model was coined VXPFC. The higher-order correlations of the model are built up as a sum of products of two-point kernels, each designed in reciprocal space to operate at both short \textit{and} long wavelengths of the density field. Through both theoretical analysis and numerical computation, the properties of phase coexistence were examined, yielding  temperature-density ($T$-$\bar\rho$) and pressure-temperature ($P$-$T$) phase diagrams incorporating the solid-liquid-vapor triple point and the liquid-vapor critical point of a pure material. We demonstrated qualitative \textit{and} quantitative agreement of the model phase diagram with experimental and atomically generated phase diagrams of aluminum over experimentally relevant ranges of temperature and density. 

We studied microstructure evolution with the proposed model, demonstrating, crucially, that it is free of numerical artifacts in the description of solid-vapor interfaces, which were found to exist in a previous vapor-forming PFC models~\cite{Emilythesis}. e then examined vapor-liquid and vapor-solid phase coexistence, and how to control their corresponding interface energy as a function of the parameters of the model's correlation function. To illustrate that our approach can be used to describe other complex lattice structures, two types of crystal lattices were studied in 2D: the triangular and the square phases.  As the approach developed here is easily extendable to 3D and a robust range of crystal structures, we expect that this modeling formalism  can serve as a valuable tool for modeling microstructure evolution in systems involving multi-phase interactions with grain boundaries, defects, voids and cracks in phenomena ranging from solidification to physical and chemical vapor deposition. 
    
    \section*{Acknowledgements}
    D. C. thanks Paul Jreidini and Matthew Frick for insightful discussions and valuable assistance. N. P. acknowledges the Natural Science and Engineering Research Council (NSERC) of Canada, Canada Research Chairs (CRC) Program for funding and Calcul Québec for computing resources.
    
    
    \appendix
    \section{Previous single-field PFC models for solid-liquid-vapor systems}\label{sec:singlefield}

    Here, we show how to recover three recent vapor-forming PFC models, following a single-field approach, discussed in the introduction. The first is that of Kocher and co-workers~\cite{kocher2015}, which can be written as
\begin{align}
    \dfrac{\Delta F}{k_B T\rho_0 a^d} =&\;
    {\int}{d\br} \big[
    \tfrac{1}{2} n^2(\mathbf{r}) 
    -\tfrac{1}{6} n^3(\mathbf{r})
    +\tfrac{1}{12} n^4(\mathbf{r})
    \nonumber\\&
    -\tfrac{1}{2} n(\br)\int 
    d\br' C_2(|\br-\br'|)n(\br')
    \nonumber\\&
    +\tfrac{1}{3}(a\Delta B+b)n(\mathbf{r})
    n_{mf}^2(\br)
    +\tfrac{1}{4}c n(\mathbf{r}) n_{mf}^3(\br)\big] \,,
    \label{eq:kochermodel}
\end{align}
where $n_{mf}(\mathbf{r})=\int d\br' \chi(\br-\br')n(\br')$ is defined as the ``density mean-field''. The bulk compressibility and the strength of the anisotropy in the periodic phase is controlled by $B^x$, while $B^\ell$ corresponds to the inverse compressibility of the liquid phase, such that $\Delta B=B^\ell-B^x$ acts as an effective temperature parameter. This model considers the original two-point correlation function proposed by Elder \textit{et al.}~\cite{elder2002,Elder2004,elder2007}, \textit{i.e.} $\hat{C}_2(\bq)=1-\Delta B -B^x(|\bq|^2-1)^2$, in Fourier space, which produces triangular (BCC) solid phase in 2D (3D). The terms coupling the density to its mean-field result from incorporating higher-order correlations, which are defined as smoothing kernels designed to operate only on the long wavelength of the density field as $\hat{\chi}(\bq)=\exp[-|\bq|^2/2\lambda^2]$. This approach was the starting point of this class of vapor PFC models that consider effective higher-order correlations to control the liquid and vapor phases. It can also be interpreted as a particular case of the model in Sec.~\ref{sec:model} for $\eta_{(2)}=\{C_2\ast n\}(\br)$, $\eta_{(3)}(\br)=\eta_{(4)}(\br)=n_{mf}(\br)$, and this suitable choice of parameters:  $C_0=-1$, $D_0=1$, $D_1=0$, $D_2=-2(a\Delta B +b)$, $E_0=-2$, $B_0=E_1=E_2=0$, and $E_3=-6c$. We also note that there is another possible choice of correlation kernels and parameters, which is found by expanding the $\eta$ functions following Eq.\eqref{eq:kernelexp}; this could be useful when fitting self-interaction terms independently. Building upon the approach in Eq.\eqref{eq:kochermodel}, a follow-up work by Kocher \textit{et al.} also considered an expansion of the PFC free energy about a Van der Waals fluid, which led to a new model shown to be robust enough to match phase diagrams rather well quantitatively~\cite{Kocherthesis}. Jreidini \textit{et al.} considered a more generalized version of Kocher's model, in which all parameters are temperature dependent (up to second order), and now the two-point direct correlation is designed as in XPFC models, with Gaussian peaks~\cite{jreidini2021,jreidini2022}:
\begin{align}
    \dfrac{\Delta F}{k_B T\rho_0 a^d} =&\;
    {\int}{d\br} \bigg[
    \tfrac{1}{2} p_2(\tau)n^2(\mathbf{r}) 
    +\tfrac{1}{3} p_3(\tau)n^3(\mathbf{r})
    \nonumber\\&
    +\tfrac{1}{4} p_4(\tau) n^4(\mathbf{r})
    +\tfrac{1}{2}q_2(\tau)
    n(\br_1)n_{mf}(\mathbf{r})
    \nonumber\\&
    +\tfrac{1}{3}q_3(\tau)
    n(\br_1)n_{mf}^2(\mathbf{r})
    +\tfrac{1}{4}q_4(\tau)
    n(\br_1)n_{mf}^3(\mathbf{r})\bigg]
    \nonumber\\&
    -\dfrac{1}{2!}{\int}{d\br}\;n(\br)\int 
    d\br' C_2(|\br-\br'|)n(\br')\,,
    \label{eq:jreidinimodel}
\end{align}
which is similar to the model in Eq.~\eqref{eq:vxpfcmodel2} for the following choice of parameters: $C_0=-p_2(\tau)$, $D_0=-p_3(\tau)$, $E_0=-2p_4(\tau)$, $D_2=-2q_3(\tau)$, $E_3=-6q_4(\tau)$, $B_0=D_1=E_1=E_2=0$, $\eta_{(2)}=-[q_2(\tau)n_{mf}+\int d\br'C_2(|\br-\br'|)n(\br')]$, and $\eta_{(3)}(\br)=\eta_{(4)}(\br)=n_{mf}(\br)$. Additionally, the mean field density is defined similarly to the one in Eq.~\eqref{eq:kochermodel}. The temperature dependence of the above parameters was chosen to be polynomial expansions up to $\mathcal{O}(\tau^2)$. Wang and co-workers developed a formalism in~\cite{wang2018angle} from which they proposed a minimal PFC model for including the vapor phase~\cite{wang2020minimal,wang2022control}. For achieving the latter, they incorporated
higher-order terms through three- and four-point correlations, expressed in real space as an expansion in gradients of the density~\cite{wang2018angle}. Their minimal vapor-forming PFC model can be written as
\begin{align}
    \dfrac{\Delta F}{k_B T\rho_0 a^d} =&
        {\int}{d\br} 
        \bigg\{
        -B_0 n(\br)
    \nonumber\\&
    -\dfrac{1}{2}n(\br)\int{d\br'}\,
        (C_0 + C_2 \nabla^2 + C_4 \nabla^4 
        \nonumber\\&
        + C_6 \nabla^6)
        \delta(\br-\br')
        \, 
        n(\br')
    \nonumber\\&
    - \dfrac{1}{3!}\bigg[D_0 n^3(\br) + D_{11} n^2(\br)\nabla^2n(\br)
    \bigg]
    \nonumber\\&
    - \dfrac{1}{4!}\bigg[E_0 n^4(\br) 
    + E_{1122} n^2(\br)[\nabla^2n(\br)]^2
    \bigg]
    \bigg\} \,,
    \label{eq:zfmodel}
\end{align}
which can be achieved by considering our model in Eq.~\eqref{eq:vxpfcmodel2}, defining the correlation kernels in real space as $\tilde{C}_2(\br-\br') = (C_{2}\nabla^2+C_{4}\nabla^4+C_{6}\nabla^6\,)\,\delta(\br-\br')$, $\tilde{C}_3(\br-\br') = \nabla^2\delta(\br-\br')$, $\tilde{C}_4(\br-\br') = \nabla^2\delta(\br-\br')$, and choosing parameters, such that:  $D_1=D_{11}$, $D_2=0$, $E_1=0$, $E_2=E_{1122}$, and $E_3=0$.

    \section{Linear stability analysis and structure factor of uniform states}\label{sec:strucfactor}

    The general expression for the structure factor (or rather, the Ursell function) can be derived for the VXPFC model by performing a linear stability analysis of the dynamical equation in Eq.~\eqref{eq:dynamicalVXPFC2} around a uniform state ($\aven$), following the similar approach as in~\cite{wang2020minimal}. The evolution equation for the density field is cast in the form $\partial_t n(\br)=G[n(\br,t)]+\xi(\br,t)$, where $G=M\nabla^2[\delta\mathcal{F}/\delta n(\br)]$, and we have introduced a mobility constant $M$, for generality. The noise $\xi(\br,t)$ is a random function with zero average value $\langle\xi(\br,t)\rangle=0$ with a characteristic timescale much smaller than that of $\delta n(\br,t)$ and its time autocorrelation function can be approximated by a $\delta$-function, such that $\langle\xi(\br,t)\xi(\br',t')\rangle=-N_a^2\nabla_{\br'}^2\delta(\br-\br')\delta(t-t')$, where $N_a^2=2Mk_BT$ is the amplitude of the noise. The motion of the nonlinear system is assumed to be in the vicinity of that from an arbitrary uniform state ($\bar{n}$), such that the density field can be decomposed as $n(\br,t)=\bar{n}+\delta n(\br,t)$, where $\delta n(\br,t)$ represents a small perturbation. Then, the right-hand side of the dynamical equation can be Taylor expanded to first order as,
\begin{align*}
    G[\bar{n}+\delta n(\br,t)] =
    G[\bar{n}] + \dfrac{\delta G}{\delta n}\bigg\vert_{n\,=\,\bar{n}}\delta n(\br,t) \,.
\end{align*}
Since $G[\delta n(\br,t)]=G[\bar{n}+\delta n(\br,t)]-G[\bar{n}]$, we have
\begin{align}
    \partial_t\delta n(\br,t)
    =-\Gamma(\bar{n},\br,\br')\delta n(\br,t) +\xi(\br,t) \,,
    \label{eq:dyneqsmall}
\end{align}
where $\Gamma(\bar{n},\br,\br')=-M\nabla^2[\delta^2\mathcal{F}/\delta n(\br)\delta n(\br')]_{n\,=\,\bar{n}}$ is the linear operator acting on $\delta n(\br,t)$. In Fourier space, Eq.~\eqref{eq:dyneqsmall} reads
\begin{align}
    \partial_t\delta\hat{n}(\bq,t)
    =-\gamma_\bq\delta\hat{n}(\bq,t) +\hat{\xi}(\bq,t)\,,
    \label{eq:linearode}
\end{align}
where $\langle\hat{\xi}(\bq,t)\hat{\xi}(\bq',t')\rangle=2Mk_BT |\bq|^2\delta(\bq-\bq')\delta(t-t')$,
and for which the solution is
\begin{align}
    \delta\hat{n}(\bq,t) =
    e^{-\gamma_\bq t} \delta\hat{n}(\bq,0)
    + e^{-\gamma_\bq t}
    \int_0^t dt' e^{\gamma_\bq t'}\hat{\xi}(\bq,t') \,.
    \label{eq_ode_solution}
\end{align}
Here, $\delta\hat{n}(\bq,t)$ and $\gamma_\bq$ are the Fourier transforms of $\delta n(\br,t)$ and the operator $\Gamma(\bar{n},\br,\br')$, respectively. The linear growth in Eq.~\eqref{eq:linearode}, for the VXPFC model, is:
\begin{align}
    \dfrac{\gamma_\bq}{\tau|\bq|^2M} = 
     & \;
        - \big[C_0 + \hat{\tilde{C}}_2(|\bq|)\big]
        -\tfrac{1}{3!}\aven
        \big[
        6 D_0 +4 (D_1 + D_2) \hat{\tilde{C}}_3(\bq)
        \big]
        \nonumber\\&
        -\tfrac{1}{4!}\aven^2
        \big[
        12 E_0 +2 (E_1 + E_2)\hat{\tilde{C}}_4(\bq)
        + E_3\hat{\tilde{C}}_4^2(\bq)
        \big] \,.
    \label{eq_gamma_q}
\end{align}
We note that Eq.~\eqref{eq_gamma_q} also lends itself to the study of how model parameters can affect numerical stability, which can also guide the appropriate choice of grid spacing and time step in dynamical simulations. 
We define the Ursell function as $\hat{S}_{nn}(\bq,t)=\langle\delta\hat{n}(\bq,t)\delta\hat{n}(-\bq,t)\rangle=\langle|\delta\hat{n}(\bq,t)|^2\rangle$, such that, using Eq.~\eqref{eq_ode_solution}, it yields
\begin{align}
    \hat{S}_{nn}(\bq,t) = e^{-2\gamma_\bq t}\hat{S}_{nn}(\bq,0)
    + \dfrac{Mk_B T |\bq|^2}{\gamma_\bq}\bigg(1-e^{-2\gamma_\bq t}\bigg) \,.
\end{align}
For times much greater than the characteristic timescale of the problem ($t\gg\gamma_\bq^{-1}$), the system is assumed to have reached equilibrium, and the Ursell function reduces to $\hat{S}_{nn}(\bq)=\hat{S}_{nn}(\bq,t\gg\gamma_\bq^{-1})=Mk_B T |\bq|^2/\gamma_\bq$, which gives
\begin{align}
    \label{eq:ursell}
    \dfrac{\hat{S}_{nn}(\bq)}{k_B T_0} =
    \bigg\{ &- \big[C_0 + \hat{\tilde{C}}_2(|\bq|)\big]
        -\tfrac{1}{3!}\aven
        \big[
        6 D_0 
        \nonumber\\&
        +4 (D_1 + D_2) \hat{\tilde{C}}_3(\bq)
        \big]
        -\tfrac{1}{4!}\aven^2
        \big[
        12 E_0 
        \nonumber\\&
        +2 (E_1 + E_2)\hat{\tilde{C}}_4(\bq)
        + E_3\hat{\tilde{C}}_4^2(\bq)
        \big]
    \bigg\}^{-1} \,,
\end{align}
using Eq.~\eqref{eq_gamma_q}. In the continuum limit, the structure factor of a uniform fluid reads $s(\bq)=\hat{S}_{nn}(\bq)+\aven (2\pi)^3\delta(\bq)$, which corresponds to the Ursell function away from $|\bq|=0$. At the uniform liquid reference density ($\aven=0$), Eq.~\eqref{eq:ursell} only has contributions from the direct two-point correlation, which then serves as a ``template'' for solid crystal growth.

    \section{Computational methodology}\label{sec:compmethod}

    We utilize a pseudo-spectral semi-implicit scheme to numerically solve Eq.~\eqref{eq:dynamicalVXPFC2}, following the approach outlined in~\cite{vitral2019role}, employed for modeling the smetic phase of liquid crystals using phase field methods. Linear terms are computed in Fourier space, while nonlinear terms are computed in real space. This choice avoids the computational overhead associated with Fourier mode convolutions, thereby reducing computational costs. Additionally, we employ a second-order accurate scheme in time. Our fast Fourier transform (FFT)-based code solves the evolution equation for the order parameter through an in-house developed {C/C++} code, which relies on the FFTW library~\cite{fftw,frigo2005design} and standard MPI libraries for parallelization.
The equation of motion for the conserved 
density field, $n(\br,t)$, can be rewritten as
$\partial_t n(\br,t) = \mathcal{L}[n(\br,t)] + \mathcal{N}[n(\br,t)]$
such that in reciprocal space the dynamics for the Fourier coefficients become ${\partial_t\hat{n}_\bq} = \hat{\mathcal{L}}_\bq + \hat{\mathcal{N}}_\bq\,$,
where $\bq$ denotes the corresponding wavenumber. The Fourier 
transform of the linear terms, $\hat{\mathcal{L}}_\bq$,
and of the nonlinear terms, $\hat{\mathcal{N}}_\bq$, are respectively defined as
\begin{align}
	\hat{\mathcal{L}}_\bq := & 
	-\tau|\bq|^2\left[-C_0-\hat{\tilde{C}}_2(\bq)\right]\hat{n}_\bq \,,
	\nonumber\\[0ex]
	\hat{\mathcal{N}}_\bq := & -\tau|\bq|^2
    \bigg\{
    -\dfrac{1}{3!}\big[
	3D_0n^2 \nonumber
    \\\nonumber&
    + D_1 \big(2n\eta_{(3)} + \tilde{C}_3\ast n^2\big)
	\\\nonumber&
	+ D_2 \big(\eta_{(3)}^2 + 2\tilde{C}_3\ast\{n\eta_{(3)}\}\big)
	\big] 
    \\\nonumber&
    - \dfrac{1}{4!}\big[
	4 E_0 n^3 
    +E_1\big(3n^2\eta_{(4)} + \tilde{C}_4\ast n^3\big)
    \\\nonumber&
	+E_2\big(2n\eta_{(4)}^2 + 2\tilde{C}_4\ast\{n^2\eta_{(4)}\}\big)
	\\&
	+E_3\big(\eta_{(4)}^3 + 3\tilde{C}_4\ast\{n\eta_{(4)}^2\}\big)\big]  
 + \xi(\mathbf{r},t)
    \bigg\}_\bq\,, 
\end{align}
where $\{\cdot\}_\bq$ represents the forward transform, and linear terms can be written as $\hat{\mathcal{L}}_\bq:=\omega_\bq\hat{n}_\bq$, so that we can identify $\omega_\bq$ as the linear growth rate. We also note that some terms in $\hat{\mathcal{N}}_\bq$ are computed following $\{\tilde{C}_{m}\ast f\}_\bq=\hat{\tilde{C}}_m(\bq)\hat{f}(\bq)$, where $f$ is just a generic function. A second-order semi-implicit scheme is implemented as a 
combination of the Crank-Nicolson (CN)
scheme for the linear terms with an explicit second-order
Adams-Bashforth (AB) scheme for the nonlinear terms in Fourier
space~\cite{vitral2019role}. The target scheme reads ${(\hat{n}_\bq^{m+1}-n_\bq^{m})}/{\Delta t}=
	\frac{1}{2}[
	\hat{\mathcal{L}}_\bq^{m+1}+\hat{\mathcal{L}}_\bq^{m}]
	+\frac{1}{2}[
	3\hat{\mathcal{N}}_\bq^{m}-\hat{\mathcal{N}}_\bq^{m-1}]\,$,
where the linear multiplier $\omega_\bq$
is evaluated just once before the main loop. The iteration scheme for the numerical solution at the new time step is
\begin{align}
    \hat{n}_\bq^{m+1}=
        \dfrac{
               \left(
                     1+\frac{\Delta t}{2}\omega_\bq
               \right)\hat{n}_\bq^{m}+
               \frac{\Delta t}{2}\left(3\hat{\mathcal{N}}_\bq^{m}-\hat{\mathcal{N}}_\bq^{m-1}\right)
               }{\left(
                       1-\frac{\Delta t}{2}\omega_\bq
                 \right)}
        \,,
        \label{eq:itescheme}
\end{align}
where the numerical solution ($\hat{n}_\bq^{m+1}$) is converted back to real space via inverse Fast Fourier Transform (iFFT) as: $n^{m+1}(\br)=[n_\bq^{m+1}]_\br:=\sum_{\bq\in\mathbb{Z}^d}\hat{n}_\bq^{m+1}e^{i\bq\cdot\br}$.

In this work, we use periodic boundary conditions for the density field in all simulations, and the computational domain
is $\Omega=[0,L]^2$, where $L$ is the domain side length. We define the grid spacing as $h = 2\pi/8|\bq_{10}|$, using the first lattice peak wavenumber given in Table~\ref{tab:modelparams}. This way, the resolution of the grid is $8$ points per each wavelength, which is shown to offer a satisfactory trade-off between precision and computational cost, considering the high spatial accuracy of spectral methods. The total number of nodes, or grid points, is $N$ (generally $512^2$, $1024^2$, $2048^2$, specified in each case), such that $L=N^{1/2}h$. We used a fixed time step of $\Delta t=0.01$ in all simulations. The Gaussian noise is generated in real space for a discrete mesh, and then filtered in reciprocal space by a cutoff of wavelengths shorter than the lattice constant $2\pi/|\bq_{10}|$, such that $\langle\xi_i(t)\,\xi_j(t^\prime)\rangle =(N_a^2/h^2\Delta t)\bq^2\delta_{ij}\delta_{t,t^\prime}$, for $|\bq|<|\bq_{10}|$, where the standard deviation is $N_a/(h^2\sqrt{\Delta t})$, hence the variance is $N_a^2/(h^2\Delta t)$.

\subsection*{Calculation of interfacial free energy between phases}

The interfacial energy of a system in a phase coexistence state can be calculated by subtracting the free energy of the bulk phases from the total free energy of the system. Although we consider simulations involving solid, liquid, and vapor in this work, we only compute the interfacial energy between two of these phases, i.e., phases $1$ and $2$. Based on the work of Wu and Karma~\cite{wu2007phase}, this free-energy difference is calculated by means of the dimensionless excess free energy, defined as 
\begin{align}
    \gamma_\text{excess} =\dfrac{1}{L}
    \int d\br \bigg[f -
    \bigg(
        f_1 \dfrac{n(\br)-\aven_2}{\aven_1-\aven_2}-
        f_2 \dfrac{n(\br)-\aven_1}{\aven_1-\aven_2}
    \bigg)\bigg]\,,
\end{align}
where, $L$ is the domain side length. Here, the free-energy density $f$ is the integrand of the VXPFC model, in Eq.~\eqref{eq:vxpfcmodel2}, i.e., $\Delta\mathcal{F}=\int d\br\, f(\br,n(\br))$, and $\aven_i$ and $f_i$ are the mean values of the numerically relaxed density field $n(\br)$ and free-energy densities for phases $1$ and $2$, respectively. This form is also addressed in~\cite{chan2015phase}.

%

    \bibliographystyle{apsrev4-1}

    \clearpage

\end{document}